\documentclass[prc,twocolumn,superscriptaddress]{revtex4-1}
\usepackage{epsfig}
\usepackage{graphicx}
\usepackage{palatino}
\usepackage[english]{babel}
\usepackage{hyphenat}
\usepackage{amsmath}
\usepackage{amssymb}
\usepackage{slashed}
\usepackage{epstopdf}
\usepackage{xcolor}
\usepackage{booktabs}
\definecolor{lcolor}{rgb}{0.,0.0,0.}
\definecolor{citcolor}{rgb}{0,0.,0.5}
\usepackage[breaklinks,colorlinks,urlcolor=blue,citecolor=blue,linkcolor=blue]{hyperref}
\usepackage{multirow}
\usepackage{ltablex}

\newcommand{\beq}{\begin{equation}}
\newcommand{\eeq}{\end{equation}}
\newcommand{\bea}{\begin{eqnarray}}
\newcommand{\eea}{\end{eqnarray}}
\newcommand{\dis}{\displaystyle}
\def\dd{{\rm d}}

\newcommand{\bem}{\begin{multline}}
\newcommand{\eem}{\end{multline}}
\newcommand{\beg}{\begin{gather}}
\newcommand{\eeg}{\end{gather}}

\newcommand{\nn}{\nonumber}

\newcommand{\ben}{\begin{eqnarray*}}
\newcommand{\een}{\end{eqnarray*}}

\newcommand{\eq}[1]{\begin{align}#1\end{align}}
\begin{document}
\title{{\bf Correlated wounded hot spots in proton-proton interactions  \\}}
\author{Javier L. Albacete}
\email[]{albacete@ugr.es}
\affiliation{CAFPE and Departamento de F\'isica Te\'orica y del Cosmos,  Universidad de Granada, E-18071 Campus de Fuentenueva, Granada, Spain.}
\author{Hannah Petersen}
\email[]{petersen@fias.uni-frankfurt.de}
\affiliation{Frankfurt Institute for Advanced Studies, Ruth-Moufang-Strasse 1, 60438 Frankfurt am Main, Germany.}
\affiliation{Institute for Theoretical Physics, Goethe University,Max-von-Laue-Strasse 1, 60438 Frankfurt am Main, Germany.}
\affiliation{GSI Helmholtzzentrum f{\"u}r Schwerionenforschung, Planckstr. 1, 64291 Darmstadt, Germany.}
\author{Alba Soto-Ontoso}
\email[]{ontoso@fias.uni-frankfurt.de}
\affiliation{CAFPE and Departamento de F\'isica Te\'orica y del Cosmos,  Universidad de Granada, E-18071 Campus de Fuentenueva, Granada, Spain.}
\affiliation{Frankfurt Institute for Advanced Studies, Ruth-Moufang-Strasse 1, 60438 Frankfurt am Main, Germany.}

\begin{abstract}
We investigate the effect of non-trivial spatial correlations between proton constituents, considered in this work to be gluonic hot spots, on the initial conditions of proton-proton collisions from ISR to LHC energies, i.e. $\sqrt s\!=\!52.6,7000,13000$~GeV. The inclusion of these correlations is motivated by their fundamental role in the description of a recently observed new feature of $pp$ scattering at $\sqrt s\!=\!7$~TeV, the hollowness effect. Our analysis relies on a Monte-Carlo Glauber approach including fluctuations in the hot spot positions and their entropy deposition in the transverse plane. We explore both the energy dependence and the effect of spatial correlations on the number of wounded hot spots, their spatial distribution and the eccentricities, $\varepsilon_n$, of the initial state geometry of the collision. In minimum bias collisions we find that the inclusion of short range repulsive correlations between the hot spots reduces the value of the eccentricity ($\varepsilon_2$) and the triangularity ($\varepsilon_3$). In turn, upon considering only the events with the highest entropy deposition i.e. the ultra-central ones, the probability of having larger $\varepsilon_{2,3}$ increases significantly in the correlated scenario. Finally, the eccentricities show a quite mild energy dependence.  

\end{abstract}

\maketitle
\section{Introduction}
Ultra-relativistic heavy-ion collisions at the Relativistic Heavy Ion Collider (RHIC) have provided strong indications of the formation of drops of quark-gluon-plasma (QGP) \cite{Arsene:2004fa}. One of the most important observables supporting the discovery of the QGP is the high value of the elliptic flow $v_2$. It quantifies how the initial spatial anisotropy of the nuclear overlap region is converted into a final state momentum space anisotropy via large collective pressure gradients during the evolution of the system. 
The Large Hadron Collider (LHC) has confirmed the appearance of a collective flow in the expanding fireball at even higher energies $\sqrt{s_{NN}}\!=\!2.76,5.02$~TeV \cite{Aamodt:2010pa,Adam:2016izf}. Originally, elementary collisions such as proton-proton ($pp$) or proton-nucleus ($pA$) were supposed to provide the binary collisions/cold nuclear matter baseline for heavy ion collisions. 
However, the analyses of particle correlations in very high multiplicity $pp$ collisions at the LHC at $\sqrt s\!=\!7$~TeV have revealed striking similarities to the $AA$ case. Suggestive signals of collective behavior such as non-negligible elliptic flow and long-range azimuthal correlations, the \textit{ridge}, have been measured \cite{Khachatryan:2010gv}. Recently, similar flow-like features have also been observed in the Run II data of the LHC at $\sqrt s\!=\!13$~TeV \cite{Aad:2015gqa,Khachatryan:2016txc}. It is worth to mention that non-zero values of $v_2$ and $v_3$ for inclusive charge particles for low multiplicity events have been recently reported \cite{Khachatryan:2016txc}

The non-zero value of $v_2$ in $pp$ collisions have lead to intense theoretical interest on the initial geometry in such systems. Although the initial state dynamics may be responsible for this observation, hydrodynamical evolution \cite{Weller:2017tsr} or a combination of both cannot be discarded \cite{Schlichting:2016kjw}. Focusing on the initial state geometry in $pp$ collisions, so far it has  been parametrized mainly in very simplistic ways: black-disk protons or a Gaussian density distribution. These ideas are being replaced by much more sophisticated models that take into account subnucleonic degrees of freedom and both density and geometrical fluctuations \cite{Mantysaari:2016ykx,Schlichting:2014ipa,Loizides:2016djv,Bozek:2016kpf,Welsh:2016siu,Dumitru:2012yr,Miller:2003kd,Schenke:2012fw,Gronqvist:2016hym,Avsar:2010rf,Mantysaari:2017cni}. While current initial condition models differ in many aspects all assume that the subnucleonic components of each colliding proton are completely independent from each other. 

On the other hand, the analyses of the $pp$ elastic differential cross section data from the TOTEM experiment at $\sqrt s\!=\!7$~TeV \cite{Antchev:2011zz} are suggestive of a new and counterintuitive feature of hadronic interactions: the maximum of the inelasticity profile is reached in non-central collisions \cite{Dremin:2015ujt,Alkin:2014rfa,Troshin:2016frs,Dremin:2016ugi}. This phenomenon, not observed before at lower energies, has been referred to as \textit{hollowness} effect in the literature \cite{Arriola:2016bxa}. The physical interpretation is that peripheral collisions are more effective producing new particles, i.e. are more inelastic, than head-on ones. A microscopic realization of the hollowness effect based on a geometrical picture has been offered in \cite{Albacete2017149}. In this work, the proton is envisaged as a system of 3 hot spots, i.e. the gluon clouds that surround the valence quarks, whose positions are subject to non-trivial spatial correlations and whose radii grow with increasing collision energy. The elastic scattering amplitude is then computed using the Glauber multiple scattering theory. Within this model, the dynamical mechanism underlying the onset of the hollowness effect is the transverse diffusion of the hot spot radius with increasing collision energy. Furthermore, the hollowness effect cannot be described in terms of uncorrelated proton structures \cite{Arriola:2016bxa}. The main goal of this paper is to explore further consequences of subnucleonic spatial correlations on the properties of the initial state in $pp$ collisions. For that purpose, we have used a Monte-Carlo Glauber approach with fluctuations in the hot spot positions in the transverse plane and in their entropy deposition. The origin of the fluctuations is intimately related with the quantum mechanical nature of the system and their importance inside the proton has been pointed out in different contexts \cite{Dumitru:2012yr,Mantysaari:2016ykx,Giacalone:2017uqx}. 

Albeit the debate on the necessity of spatial correlations inside the proton is very timely it has been previously discussed and analyzed in the case of nuclei \cite{Alvioli:2009ab,Blaizot:2014wba}. In particular, it has been shown that their effect on the initial condition of heavy-ion collisions leads to a non-negligible reduction of the eccentricities \cite{Denicol:2014ywa}. The sensitivity of smaller systems such as $pp$ collisions to the fine details of the initial geometry is expected to be larger than in the nucleus-nucleus case. As we shall explain thoroughly in the following sections we find sizable differences in the calculation of properties of the initial state such as the eccentricities arising from the inclusion of short-range repulsive correlations. In line with our expectations the net effect of considering correlated constituents depend on the \textit{centrality} of the collision.  In this work we characterize the centrality of an event by its deposited entropy that is tightly related with the event multiplicity. In other words, the more entropy is deposited by the wounded hot spots, the more central the event. When no cuts on the entropy deposition are applied i.e. considering minimum bias events the values of the ellipticity and triangularity obtained with correlated constituents are systematically smaller than those obtained within the uncorrelated scenario. Moreover, motivated by the phenomenological interest in very high-multiplicity $pp$ collisions we have computed the probability distributions $\mathcal P(\varepsilon_{2,(3)})$ for the $0\!-\!1$\% centrality class. After selecting the most entropic events we find that the presence of correlations boost the probability of having larger values of $\varepsilon_2$ and $\varepsilon_3$, in contrast to the minimum bias case.

This paper is organized as follows. In the next section, a detailed description of the building blocks of our Monte-Carlo Glauber calculation is presented. Next, in Section~\ref{ress} the influence of the repulsive correlations on properties of the initial state is studied. In particular, we focus our attention on the mean number of wounded hot spots, their spatial distribution in the transverse plane and, especially, on the values of $\varepsilon_2$ and $\varepsilon_3$ for different centrality classes. All these observables are affected by the inclusion of repulsive correlations. Finally, we show the energy dependence of the spatial eccentricities from ISR ($\sqrt s\!=\!52.6$~GeV) to LHC ($\sqrt s\!=\!7,13$~TeV) energies.
\section{Set Up}
\label{setup}
To compute the spatial eccentricities and have access to event-by-event fluctuations we have developed a Monte-Carlo Glauber event generator inspired by \cite{Broniowski:2007nz,Alver:2008aq} adapted to $pp$ collisions. Its main ingredients based on \cite{Albacete2017149} are revisited in this Section. The study of the effect of spatial correlations in other initial state models such as the IP-Glasma \cite{Schenke:2012wb} or the MC-rcBK \cite{Albacete:2011fw,Albacete:2012xq} is left for future work.

First of all, the impact parameter of the collision is chosen randomly
from the distribution
\eq{
\dd N_{\rm{ev}}/\dd b\propto b
}
up to $b_{\rm max}\!=\!2$~fm $\!\gtrsim\!2R_{p}$. In our picture, the impact parameter is the distance between the centers of the two protons in the $x$-direction. Thus, the centers of the colliding protons are located at $(x,y)\!=\!(-b/2,0)$ and $(b/2,0)$. Furthermore, the $z$-component will be neglected in the whole calculation i.e. we work exclusively in the transverse plane. We have checked that beyond $b\!=\!2$~fm the number of events with at least one collision is negligible.

We describe $pp$ interactions as a collision of two systems, each one composed of three hot spots. The positions of the three hot spots ($\vec{s_i}$) in each proton are sampled from the distribution \cite{Albacete2017149}
\eq{
D(\vec s_{1},\vec s_{2},\vec s_{3})&=C \dis\prod_{i=1}^3e^{- s_i^2/R^2}\delta^{(2)}(\vec{s}_1+\vec{s}_2+\vec{s}_3)
\times \nn \\
&\dis\prod_{\substack{{i<j}\\{i,j=1}}}^3\left(1-e^{-\mu\vert\vec{s}_i-\vec{s}_j\vert^2/R^2}\right).
\label{corr}
}
The constant $C$ ensures that the probability distribution is normalised to unity: $\int\lbrace\dd^2s_i\rbrace D(\lbrace s_i\rbrace)\!=\!1$. The next term corresponds to the product of three uncorrelated probability distributions for a single hot spot, where $R$ is the average radius. The $\delta$-function in Eq.~\ref{corr} guarantees that the hot spots system is described with respect to the proton centre of mass. The last term implements repulsive short-range correlations between all pairs of hot spots controlled by an effective repulsive core $r_c^2\equiv R^2/\mu$. In the limit $\mu\to\infty$, or equivalently $r_c\!=\!0$, we recover the uncorrelated case. The inclusion of this repulsive distance is the main novelty of this work with respect to others in the literature, where the subnucleonic structure was already considered \cite{Loizides:2016djv,Bozek:2016kpf,Welsh:2016siu}. It should be noted that we do not impose any kind of minimum distance between the hot spots as it is done in other works in the literature to mimic the short range correlations \cite{Broniowski:2007nz}. We generate the polar coordinates of the three hot spots, that are next easily converted into Cartesian ones, sampling $D(\vec s_{1},\vec s_{2},\vec s_{3})$ for each proton.

Once the hot spots of projectile and target are located in the transverse plane, the probability of two hot spots to collide is sampled from the inelasticity density
\eq{
G_{\rm{in}}(d)&=2e^{-d^2/2R_{hs}^2}-(1+\rho_{hs}^2)e^{-d^2/R_{hs}^2}
\label{gin}
}
where $d$ is the transverse distance between a pair of hot spots with radius $R_{hs}$, and $\rho_{hs}$ is the ratio of real and imaginary parts of the hot spot-hot spot scattering amplitude. This collision probability results from a Gaussian parametrization of the elastic scattering amplitude \cite{Albacete2017149}. We evaluate this probability for all pairs of hot spots and refer to them as \textit{wounded} \cite{Bialas:1977en,Bialas:1976ed}, if they have suffered at least one collision. Thus, the maximum number of wounded hot spots, $N_w$, in one event is 6. Another possibility that has been studied in the literature is to consider the number of binary collisions instead of the wounded hot spots scenario or a combination of both. We have tested that our main conclusions are not affected by this choice and take the wounded hot spot approach for simplicity. For each event, we keep track of the position of each wounded hot spot, ($x_w,y_w$), for later usage in the calculation of spatial distributions, eccentricities or any other quantity of interest. 

Our model has 4 free parameters $\lbrace R_{hs}$, $R$, $r_c$, $\rho_{hs}\rbrace$. For a given value of $r_c$ we constrain $\lbrace R_{hs}$, $R$, $\rho_{hs}\rbrace$ to reproduce the measured values of the total $pp$ cross section ($\sigma_{\rm tot}$) and the ratio of real and imaginary parts of the scattering amplitude ($\rho$) at each collision energy \cite{Amaldi:1979kd,Antchev:2011zz}. For the LHC at $\sqrt s\!=\!13$~TeV no experimental measurements of these quantities are yet available so we rely on the extrapolated values provided by the COMPETE collaboration \cite{Cudell:2002xe}. Upon imposing these constraints we ensure that our results are phenomenologically compatible.
Regarding the correlation structure of the hot spots, Eq.~\ref{corr}, we have considered two extreme scenarios: the uncorrelated case labeled as $r_c\!=\!0$ and a repulsive core of $0.4$~fm labeled as $r_c\!=\!0.4$. Being the main goal of this work to explore the net effect of correlations we have considered a third situation, $r_c\!=\!0,nc$, in which we set the repulsive distance to 0 but choose the values of $\lbrace R_{hs}$, $R$, $\rho_{hs}\rbrace$ as in the $r_c\!=\!0.4$ case, not reproducing though the experimental values of $\sigma_{\rm tot}$ and $\rho$. The main reason to consider this additional possibility is that differences between the results of $r_c\!=\!0.4$ and $r_c\!=\!0,nc$ are then only attributable to the presence of short-range repulsive correlations as the rest of the parameters remain identical. However for the same values of $\lbrace R_{hs}$, $R$, $r_c$, $\rho_{hs}\rbrace$ the hot spots of the correlated distribution have a larger mean transverse position, $\langle s_1 \rangle$, defined as

\eq{
\langle s_1 \rangle&=\dis\int s_1\dd \vec{s_1}\dd\vec{s_2}\dd \vec{s_3} D(\vec s_{1},\vec s_{2},\vec s_{3})
\label{<s1>}
}

where $D(\vec s_{1},\vec s_{2},\vec s_{3})$ is given by Eq.~\ref{corr}, than in the uncorrelated case. In order to avoid this artificial swelling we have included one last scenario, labeled as "$\langle s_1 \rangle$ fixed", in which $\lbrace R_{hs}$, $r_c$, $\rho_{hs}\rbrace$ are the same as in the $r_c\!=\!0.4$ case but $R$ is chosen to reproduce the $\langle s_1 \rangle$ of the correlated distribution. The values of $R$ for this case are shown in Table~\ref{param}. It should be noted that we compute $\sigma_{\rm tot}$ and $\rho$ using the Glauber multiple scattering framework as described in \cite{Albacete2017149}. Then, the values of the parameters of our model that fulfill the phenomenological conditions are not unique but rather conform a whole region of the parameter space. In Table~\ref{param} we show representative values of those allowed regions. For the repulsive distance we have also chosen an intermediate value of the ones considered in \cite{Albacete2017149}. In addition, the parameters are compatible with the hollowness effect at the pair of LHC energies considered in the $r_c\!=\!0.4$ case.
\begin{table}
\begin{center}
\begin{tabular}{c|*{6}{c|}}
  & \multicolumn{2}{|c|} {$\bf{r_c\!=\!0.4}$~{\bf fm}} & \multicolumn{2}{|c|}{$\bf{r_c\!=\!0}$} 
  & \multicolumn{1}{|c|} {$\bf{\langle s_1 \rangle}$ \bf fixed} \\ \toprule[0.5mm]
$\sqrt{s}$~[GeV] & $R_{hs}$~[fm]&$R_p$~[fm] & $R_{hs}$~[fm]&$R_p$~[fm] & $R$~[fm]  \\ \toprule[0.5mm]
52.6 & 0.19&~0.68 & 0.23&0.67& 0.84 \\ \hline
7000 & 0.3&0.75 & 0.39&0.76 & 0.83 \\ \hline
13000 & 0.32&0.8 & 0.41&0.86 &0.87
\end{tabular}
\caption{Parameters of the hot spot distribution and the inelasticity profile in Eqs.\ref{corr}-\ref{gin} for various $pp$ collision energies with ($r_c\!=\!0.4$~fm) and without ($r_c\!=\!0$) short-range repulsive correlations. $R_p$ stands for the proton radius defined as $R_p\equiv\sqrt{R^2+R_{hs}^2}$. We set $\rho_{hs}\!=\!0.1$ in all cases. On the last column, the values of $R$ for the "$\langle s_1 \rangle$ fixed" case are shown.}
\label{param}
\end{center}
\end{table}

A quantitative measurement of the initial anisotropy of the geometry in a collision is given by the spatial eccentricities that are defined as 
\eq{\varepsilon_n=\dis\frac{\sqrt{\langle\dis\sum_{i=1}^{N_w} r_{i}^{n}\cos(n\phi_i)\rangle^2+\langle\dis\sum_{i=1}^{N_w} r_{i}^{n}\sin(n\phi_i)\rangle^2}}{\langle\dis\sum_{i=1}^{N_w} r_{i}^{n}\rangle}
\label{ecc}
}
where the sum runs over all wounded hot spots. The polar coordinates ($r_i,\phi_i$) entering Eq.~\ref{ecc} are obtained from the original ones, ($x_w,y_w$), by applying two transformations. First, shift coordinates such that $(0,0)$ coincides with the center of mass of the participants system i.e. ($x_{pp}^{i}=x_{w}^i-x_w^{\rm C.o.M}$,$y_{pp}^{i}=y_{w}^i-y_{w}^{\rm C.o.M}$). Next we determine the angular orientation of the $\varepsilon_n$ plane from
\eq{\psi_n=\dis\frac{1}{n}\arctan2\left(\dis\frac{\dis\sum_{i=1}^{N_w} r^{i}_{pp}\sin(n\phi^{i}_{pp})}{\dis\sum_{i=1}^{N_w} r^{i}_{pp}\cos(n\phi^{i}_{pp})}\right)
\label{psiplane}
}
and explicitly rotate the coordinates by $\psi_n$. The shifted and rotated coordinates are the ones involved in the calculation of the eccentricities in Eq.~\ref{ecc}. In this new reference frame, often called \textit{participant plane} in the literature \cite{Qiu:2011iv,Luzum:2013yya}, $\langle x \rangle = \langle y \rangle = 0$. In our calculation $\varepsilon_n$ are defined on an event-by-event basis. Finally, $\langle \cdot \rangle$ in Eq.~\ref{ecc} denotes the average weighted by the entropy deposition. 
\begin{figure}[htb]
\begin{center}
\includegraphics[scale=0.465]{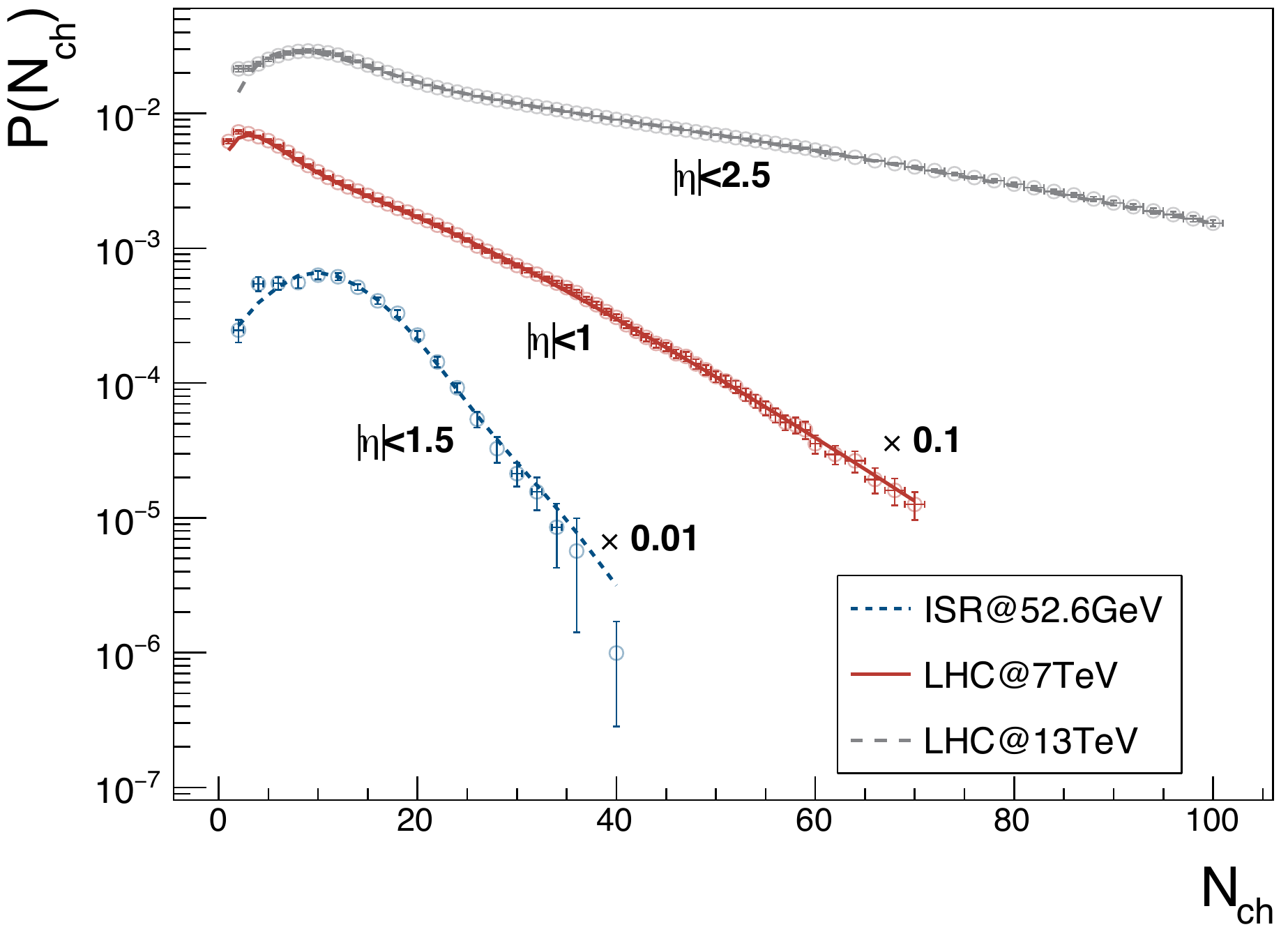} 
\end{center}
\vspace*{-0.5cm}
\caption[a]{Fit to the charged particle multiplicity distributions for different collision energies in the wounded hot spot model, Eqs.~\ref{pnch}-\ref{pnch2}. The experimental data from top to bottom is taken from: ATLAS Collaboration \cite{Aaboud:2016itf}, ALICE Collaboration \cite{Aamodt:2010pp} and ISR \cite{Breakstone:1983ns}. Note that each experiment has a different rapidity acceptance window, $\eta$, that influences the shape of the data. Also, the ALICE (red) and ISR (blue) curves are multiplied by 0.1 and 0.01 respectively.}
\label{multip}
\end{figure}
To characterize the entropy deposition we rely on a very similar approach to the one recently proposed in \cite{Bozek:2016kpf}. Essentially, entropy deposition is directly related to the number of charged particles produced in $pp$ collisions. The charged hadron probability distribution in an incoherent description of the particle production process can be written as

\eq{\mathcal{P}(N_{\rm{ch}})&=\sum_{i=2}^{N_w}\mathcal{P}_{w}(i)
\dis\sum_{n_1,n_2,\ldots,n_i}\mathcal{P}_{hw}(n_1)\mathcal{P}_{hw}(n_2)\ldots \mathcal{P}_{hw}(n_i) \nn \\
&\times\delta(N_{\rm ch}-n_1-n_2-\ldots -n_i)
\label{pnch}
}

where $\mathcal{P}_w$ is the probability distribution of $i$ hot spots to be wounded and $P_{hw}$ is the distribution of the number of hadrons produced by a single wounded hot spot. According to Eq.~\ref{pnch} each wounded hot spot contributes independently to the total charged hadron multiplicity distribution in $pp$ collisions. An important comment is in order at this point. Particle production is treated incoherently, since we assume that each hot spot contributes the same way to the multiplicity distribution independent of the number of interactions. However, it seems reasonable to think that a certain degree of coherence should be included in a realistic model for particle production. By \textit{coherence} we refer to the fact that an event where e.g. one hot spot in the projectile undergoes simultaneous scattering with three constituents in the target may not contribute in the same way to $\mathcal{P}(N_{\rm ch})$ as the incoherent superposition of three 1 vs. 1 interactions. For instance, in the IP-Glasma model the saturation scale is used as a degree of freedom to describe coherence \cite{Schenke:2012wb}. Exploring a more realistic coherent description of the charged hadron multiplicity distribution is left for future work. 

Up to this point there is still one missing element in Eq.~\ref{pnch}: the precise functional form for the hadron multiplicity distribution from each wounded hot spot $\mathcal{P}_{hw}(N_{\rm ch})$. The latest analysis of experimental data on charged hadron multiplicities by the LHC collaborations has revealed that a double negative binomial function provides a better description of the data than just a single one \cite{Adam:2015gka}. This is the choice adopted in this work to parametrize $\mathcal{P}_{hw}$

\eq{\mathcal{P}_{hw}(N_{\rm ch})&=\alpha\dis\frac{\Gamma(N_{\rm ch}+\kappa_1)\overline n_1^{N_{\rm{ch}}}\kappa_1^{\kappa_1}}{\Gamma(\kappa_1)N_{\rm ch}!(\overline n_1+\kappa_1)^{N_{\rm ch}+\kappa_1}}+\nn \\
&(1-\alpha)\dis\frac{\Gamma(N_{\rm ch}+\kappa_2)\overline n_2^{N_{\rm{ch}}}\kappa_2^{\kappa_2}}{\Gamma(\kappa_2)N_{\rm ch}!(\overline n_2+\kappa_2)^{N_{\rm ch}+\kappa_2}}
\label{pnch2}
}

where $\Gamma(x)$ is the Euler Gamma function, the averages are given by $\overline n_i$, larger $\kappa_i$ means smaller fluctuations and $\alpha$ is a mixing parameter. The parameters $\lbrace \overline n_i,\kappa_i,\alpha \rbrace$ are adjusted to reproduce the observed multiplicity distributions at all the collision energies considered in this work, independently. We achieve a good description of the data, $\chi^2/{\rm d.o.f}\sim 1.2\!-\!2$, for all cases as it is shown in Fig.~\ref{multip}. Nevertheless, there are small departures at low values of $N_{\rm ch}$ at $\sqrt s\!=\!13$~TeV and we overshoot the tail of the ISR data. We have not included in the fit the last 5 points of the experimental data of the ATLAS Collaboration at $\sqrt s\!=\!13$~TeV due to the large systematic uncertainties and their tiny contribution to the probability distribution. The results shown in Fig.~\ref{multip} correspond to the $r_c\!=\!0.4$ case. Good quality fits were found in the other scenarios as well.

\begin{figure}[h]
\begin{center}
\includegraphics[scale=0.465]{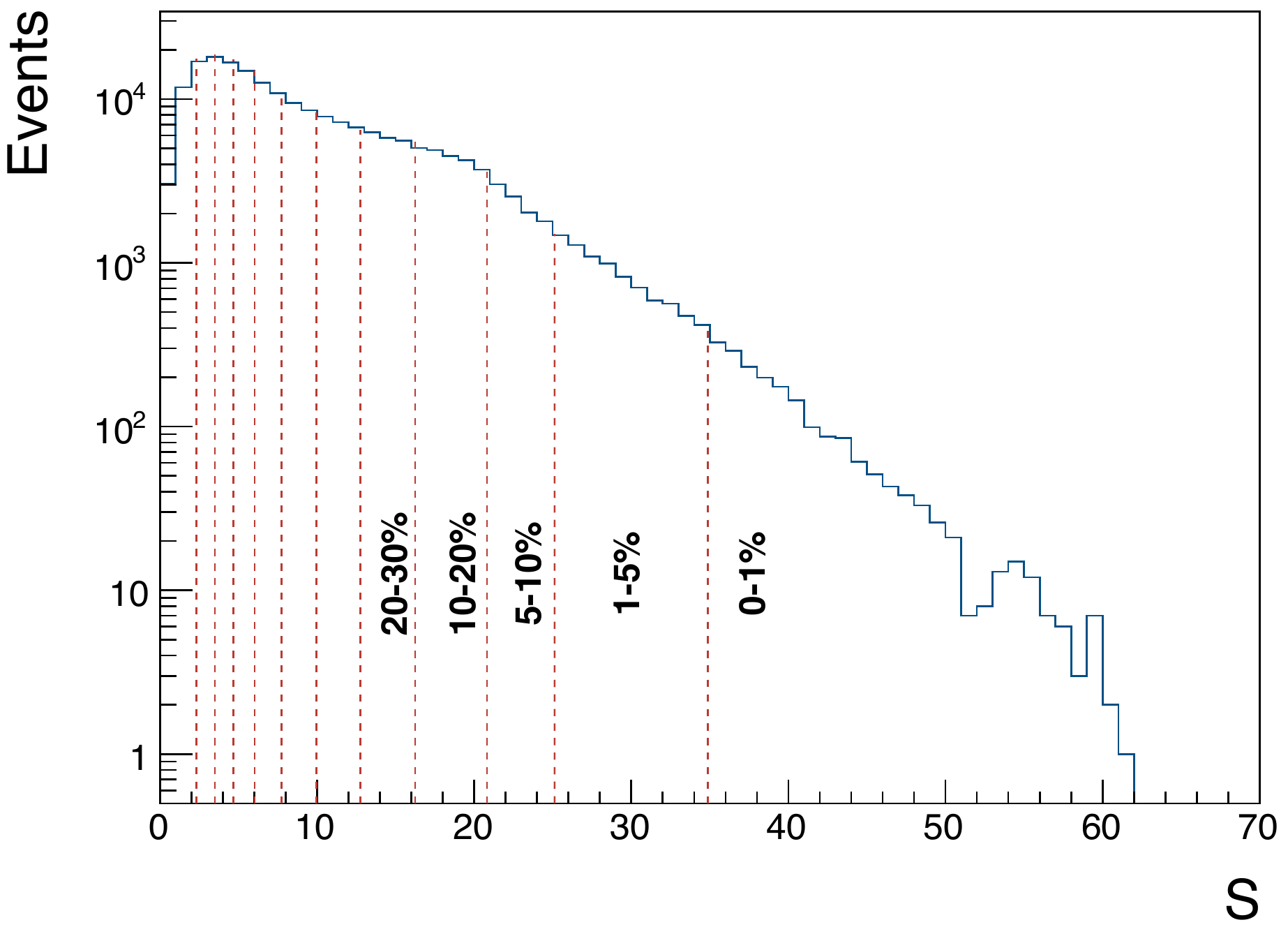} 
\end{center}
\vspace*{-0.7cm}
\caption[a]{The histogram of the integrated entropy deposition for the $r_c\!=\!0.4$~fm case at $\sqrt s\!=\!7$~TeV. Vertical red lines labelled by black numbers define centrality classes as fractions of the total number of events.}
\label{entropy_deposition}
\end{figure}

Within the current context, the main purpose of an accurate description of the charged hadron multiplicity distribution is to provide phenomenological guidance to a non-measurable quantity, the shape of the entropy distribution. In general, the negative binomial distribution can be expressed as a convolution of a Gamma and a Poisson distributions. Following the usual assumption that the particle emission is given by a Poissonian process with the mean proportional to the entropy deposited in the fluid element, the entropy distribution can be written as a double Gamma distribution:

\eq{\mathcal{P}(s_0)=&\alpha\dis\frac{s_0^{\kappa_1-1}\kappa_1^{\kappa_1}}{\Gamma(\kappa_1)\overline n_1^{\kappa_1}}\exp{(-\kappa_1 s_0/\overline n_1)}+\nn \\
&(1-\alpha)\dis\frac{s_0^{\kappa_2-1}\kappa_2^{\kappa_2}}{\Gamma(\kappa_2)\overline n_2^{\kappa_2}}\exp{(-\kappa_2 s_0/\overline n_2)}.
\label{entrop}
}
The parameters $\lbrace \overline n_i,\kappa_i,\alpha \rbrace$ in Eq.~\ref{entrop} are identical to the ones of the negative binomial distribution, Eqs.~\ref{pnch}-\ref{pnch2}, that yield a precise description of the measured multiplicity distributions, as it is depicted in Fig.~\ref{multip}. Furthermore, the entropy deposition of each wounded hot spot located at ($x_w$,$y_w$) is smeared around the center of the wounded hot spot following a Gaussian prescription

\eq{s(x,y)=s_0\dis\frac{1}{\pi R_{hs}^2}\exp\left(-\dis\frac{(x-x_w)^2+(y-y_w)^2}{R_{hs}^2}\right)
}
in order to avoid unphysical spiked entropy deposition and endowing our model with a more realistic description. In our calculation, $s_0$ fluctuates independently according to Eq.~\ref{entrop} for each wounded hot spot. In Fig.~\ref{entropy_deposition} we show the integrated entropy deposition distribution for the $r_c\!=\!0.4$~fm case at $\sqrt s\!=\!7$~TeV where $S$ is computed in each event as 
\eq{S=\dis\sum_{i=2}^{N_{w}}s_0^{i}}
and $s_0$ is given by Eq.~\ref{entrop}. We have superimposed the division of the events in centrality classes depending on their contribution to the integrated entropy.

In the following sections we present the results obtained within the framework of the Monte-Carlo Glauber model discussed above for different centrality classes: $[0\!-\!1\%], [1\!-\!5\%],[5\!-\!10\%],[10\!-\!20\%],[20\!-\!30\%]\ldots[90\!-\!100\%]$.

\section{Results}
\label{ress}
All the results presented in this Section have been obtained after generating 500k events. The averages have been performed over the number of events with at least one hot spot-hot spot collision.
\subsection{Impact of correlations}
We present our results in two different cases: all the events are selected (minimum bias) and only the events on the 0-1\% centrality class (ultra-central collisions) as defined in Fig.\ref{entropy_deposition} are considered.
\subsubsection{Minimum bias}
\begin{figure}[htb]
\begin{center}
\includegraphics[scale=0.46]{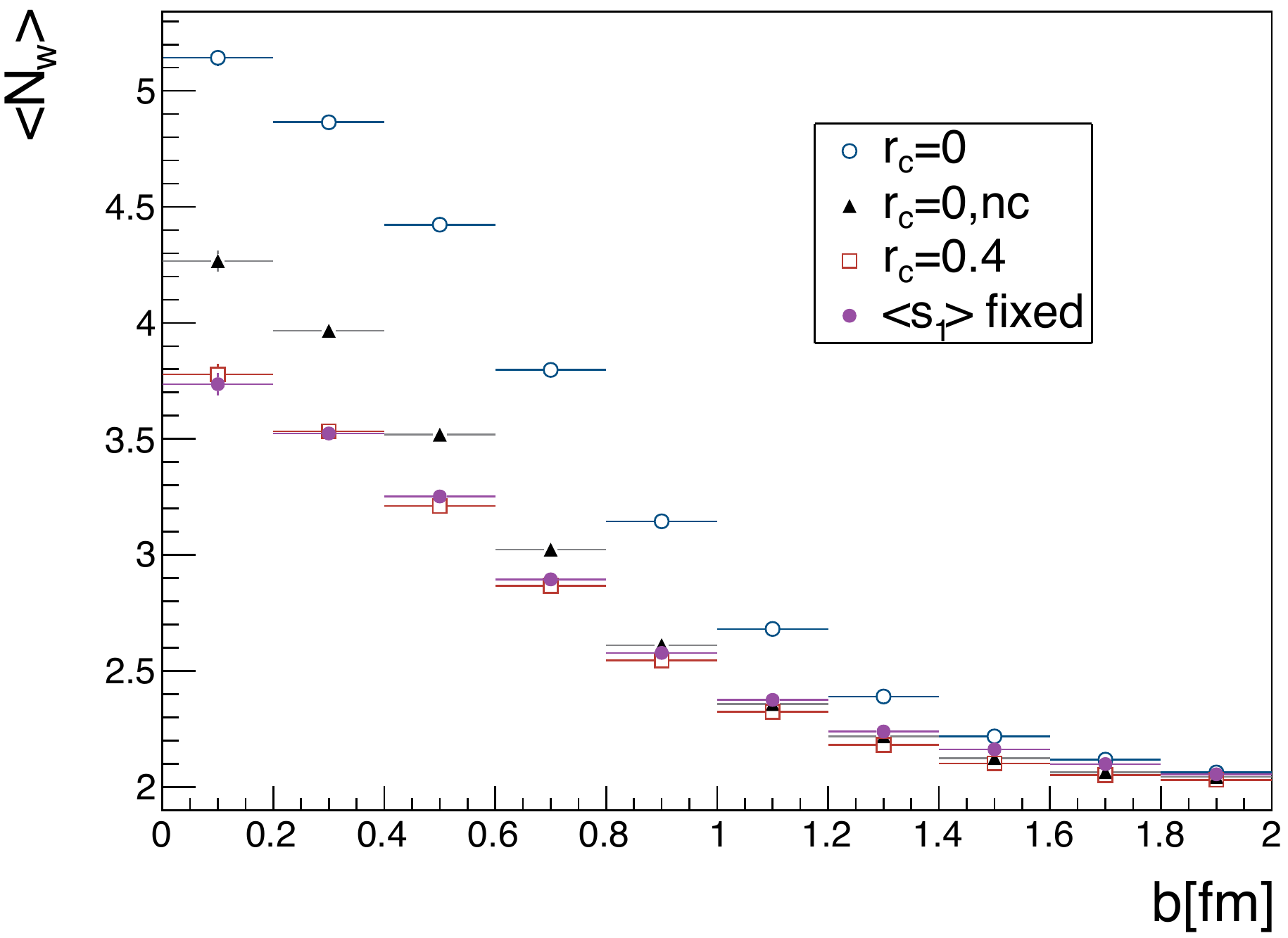} 
\end{center}
\vspace*{-0.5cm}
\caption{Average number of wounded hot spots for different impact parameter bins of the collision. The horizontal lines indicate the width of the bins.}
\label{nwounded}
\end{figure}

We begin our analysis by computing the average number of wounded hot spots in $pp$ collisions as a function of the impact parameter $b$ for the four different scenarios introduced above. The results are shown in Fig.~\ref{nwounded}. We note that the qualitative behavior of the impact parameter dependence of $\langle N_w \rangle$ is not affected by the inclusion of correlations. For instance, the number of wounded hot spots is larger in central collisions ($b\!=\!0$) than in peripheral ones, as expected. However, in central to moderately peripheral collisions, $0\!<\!b\!<\!0.8$~fm, the average number of wounded hot spots is smaller in the correlated scenario (squares vs. empty dots/triangles in Fig.~\ref{nwounded}). We have also computed the mean number of wounded hot spots in a proton-proton interaction defined as
\eq{\overline N_w=(\dis\sum_{i=1}^{N_{\rm ev}}N_w^{i})/N_{\rm ev}
\label{entropy_eq}
}
where $N_{\rm ev}$ is the total number of events with at least one collision. We find that $\overline N_w$ is slightly reduced $\sim 5\%$ in the $r_c\!=\!0.4$ case with respect to fixing $\langle s_1\rangle$. We find that a very basic element of all Monte-Carlo Glauber calculations i.e. the mean number of wounded objects, hot spots in our case, is already affected by the modification of the initial geometry of the collision as it is illustrated in Fig.~\ref{nwounded}.
\begin{figure}[h]
\includegraphics[scale=0.465]{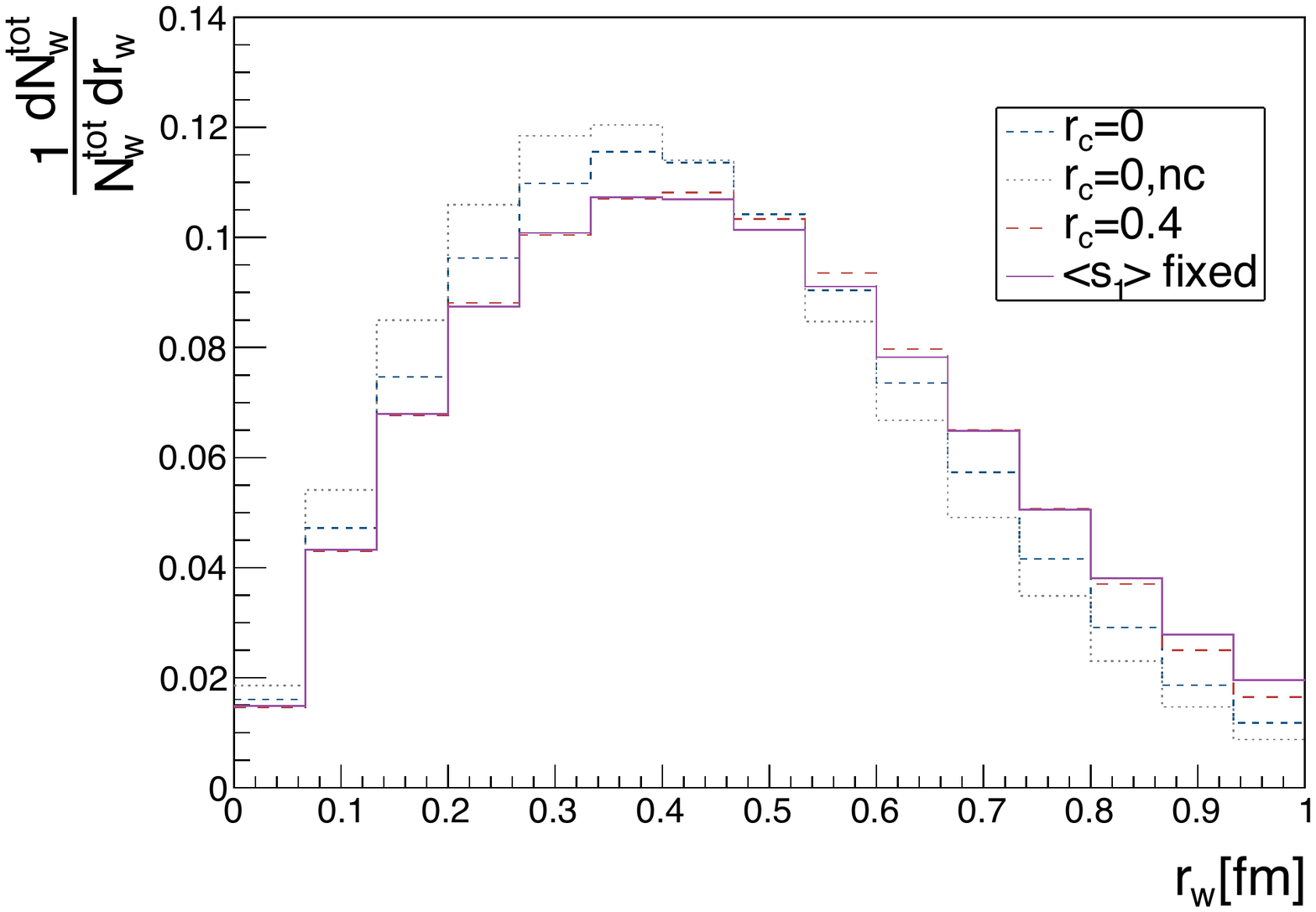} 
\vspace*{-0.5cm}
\caption[a]{Normalized radial distribution of the wounded hot spots before shifting and rotating to the participant plane.}
\label{radial}
\end{figure}
\begin{figure}[h]
\begin{center}
\includegraphics[scale=0.465]{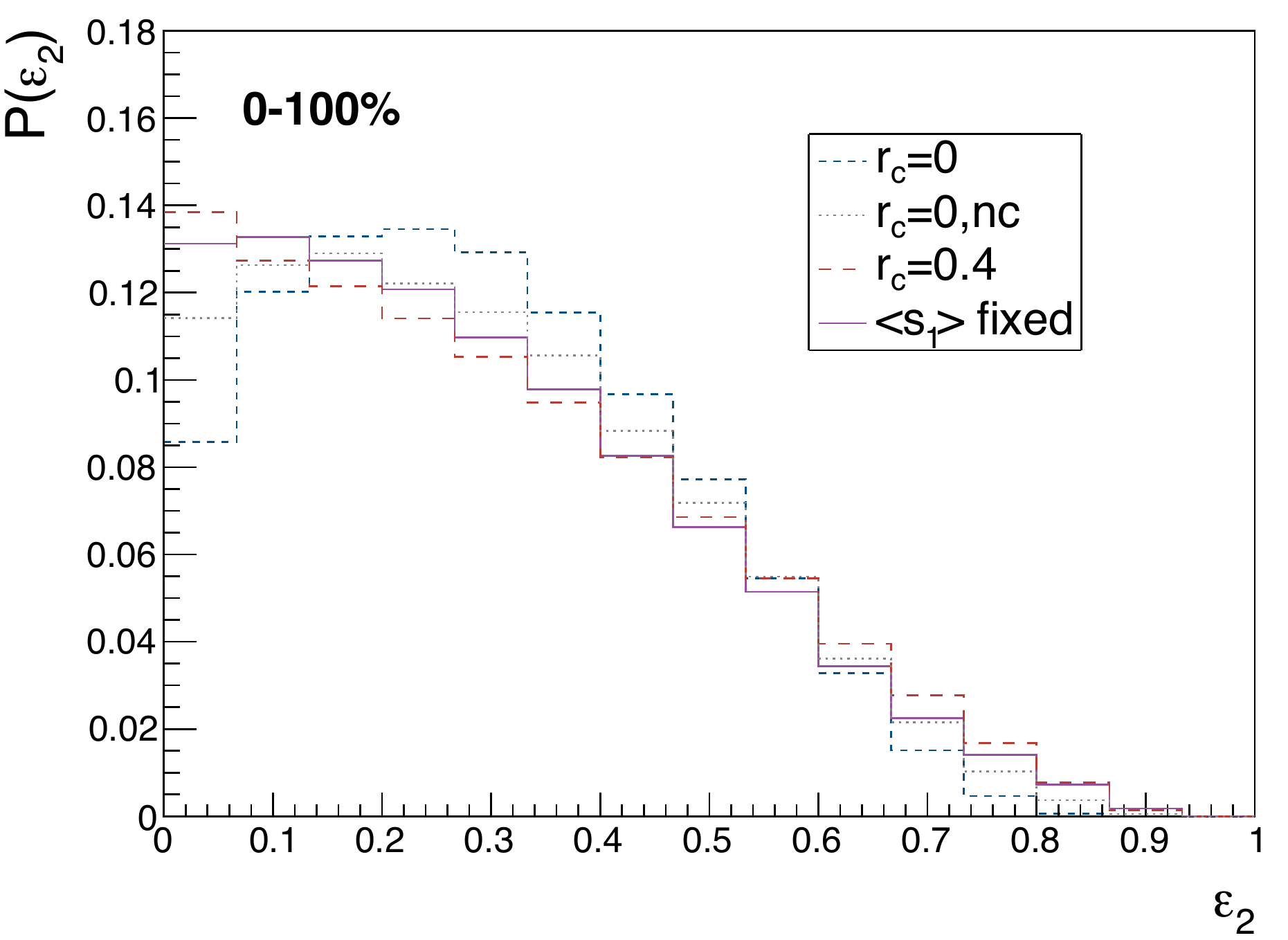} 
\end{center}
\vspace*{-0.7cm}
\caption[a]{Probability distribution of the eccentricity, $\varepsilon_2$, for $r_c\!=\!0$ (blue short-dashed line), $r_c\!=\!0.4$~fm (red long-dashed line), $r_c\!=\!0,nc$ (grey dotted line) and  $\langle s_1 \rangle$ fixed (purple solid line).}
\label{eps2}
\end{figure}

\begin{figure}[h]
\begin{center}
\includegraphics[scale=0.465]{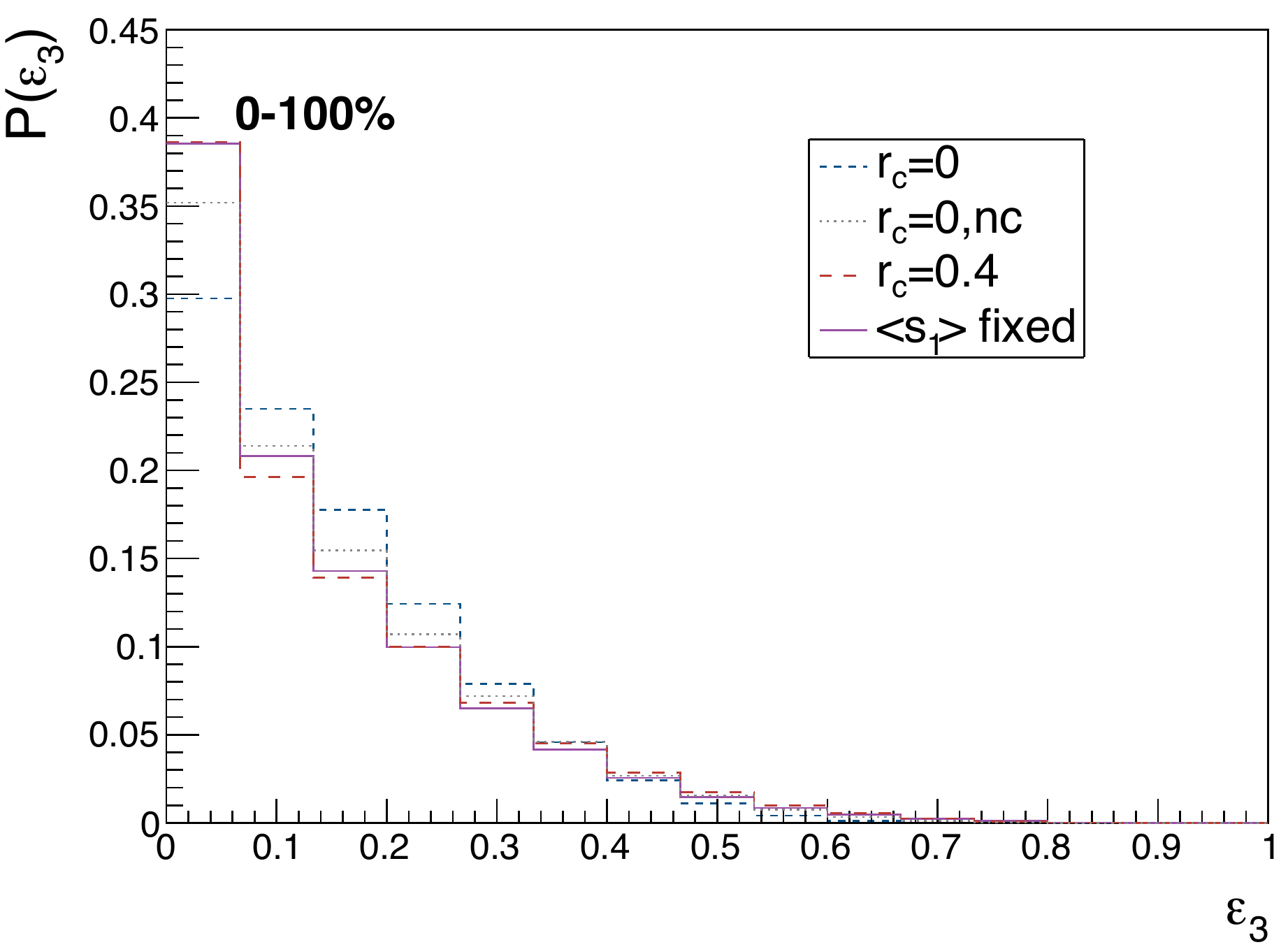} 
\end{center}
\vspace*{-0.7cm}
\caption[a]{Probability distribution of the triangularity, $\varepsilon_3$, for $r_c\!=\!0$ (blue short-dashed line), $r_c\!=\!0.4$~fm (red long-dashed line), $r_c\!=\!0,nc$ (grey dotted line) and  $\langle s_1 \rangle$ fixed (purple solid line).}
\label{eps3}
\end{figure}
Another important feature of the presence of repulsive correlations is their effect on the spatial distribution of the hot spots. In Fig.~\ref{radial} we compare the normalized radial distribution of the wounded hot spots, characterized by their polar coordinate $r_w\!=\!\sqrt{x_w^2+y_w^2}$, resulting from the uncorrelated and correlated scenarios. One sees that for $r_c\!=\!0.4$~fm, the radial distribution gets broader and its mean value is shifted to larger values than in the uncorrelated case. Therefore, a plausible interpretation in a geometrical picture is that when including repulsive correlations the probability to find wounded hot spots on the edges of the interaction region is increased.

The main result of this Section is shown in Fig.~\ref{eps2}: the eccentricity, $\varepsilon_2$, is reduced in the correlated scenario compared to the rest of the cases including the one with $\langle s_1 \rangle$ fixed. We hence conclude that the probability of having smaller values of the eccentricity in a proton-proton interaction is increased when repulsive short-range correlations are included. In essence the eccentricity is a direct measurement of the anisotropy of the interaction region between the $x$ and $y$ directions. Thus, the results presented in Fig.~\ref{eps2} suggest that the characteristic ellipsoidal shape of the interaction region between the two protons is replaced by a more round one (with smaller eccentricity) in the correlated scenario.  However, all cases exhibit a broad probability distribution of $\varepsilon_2$ due to the highly fluctuating nature of the system. 
In addition, the effect of correlations between the constituents of the proton is shown to be qualitatively the same as in the nucleus case \cite{Denicol:2014ywa} but has a stronger impact on the numerical values of $\varepsilon_2$. This is not a surprising result as smaller systems are expected to be more sensitive to the fine details of the geometry than the complex $AA$ case, where the net effect of these subtleties is washed out by the accumulation of uncorrelated nucleon-nucleon collisions. However, this study being a multiparametric one the magnitude of the eccentricity's depletion could vary depending on the values of $\lbrace R_{hs}$, $R$, $r_c$, $\rho_{hs}\rbrace$ provided that it will always decrease when including correlations among subnucleonic degrees of freedom in the proton. 

To conclude this Section we show in Fig.~\ref{eps3} the probability distribution for the triangularity, $\mathcal{P}(\varepsilon_3$). The origin of odd eccentricity moments, in our model, is not driven by the geometry of the collision but rather by the fluctuations in both the entropy deposition and the positions of the hot spots. The main reason is that our spatial distributions are symmetric with respect to the $y$-axis, $\langle y \rangle\!=\!0$, so in absence of fluctuations all the odd eccentricity moments would exactly vanish. Compared to $\varepsilon_2$, we observe that $\varepsilon_3$ is smaller and the role of spatial correlations is weaker. Nevertheless, the triangularity of the proton-proton interaction in our model shows the same qualitative behavior as the eccentricity i.e. is reduced in the correlated scenario.

\subsubsection{Ultra-central collisions}

\begin{figure}[h]
\begin{center}
\includegraphics[scale=0.465]{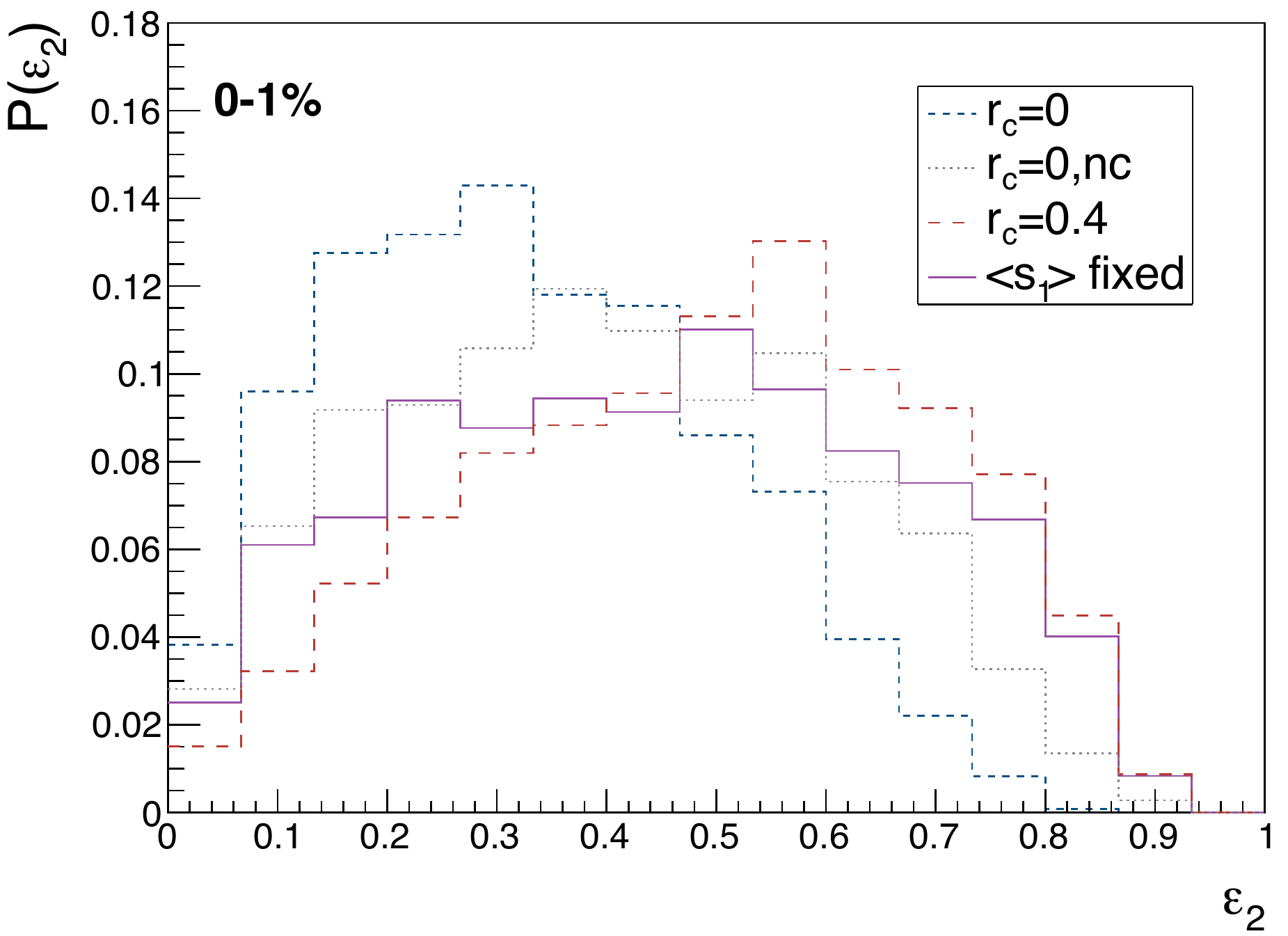} 
\end{center}
\vspace*{-0.7cm}
\caption[a]{Probability distribution of the eccentricity, $\varepsilon_2$, for $r_c\!=\!0$ (blue short-dashed line), $r_c\!=\!0.4$~fm (red long-dashed line), $r_c\!=\!0,nc$ (grey dotted line) and  $\langle s_1 \rangle$ fixed (purple solid line) after selecting the 1\% most \textit{entropic} events.}
\label{eps2_01}
\end{figure}

\begin{figure}[h]
\begin{center}
\includegraphics[scale=0.465]{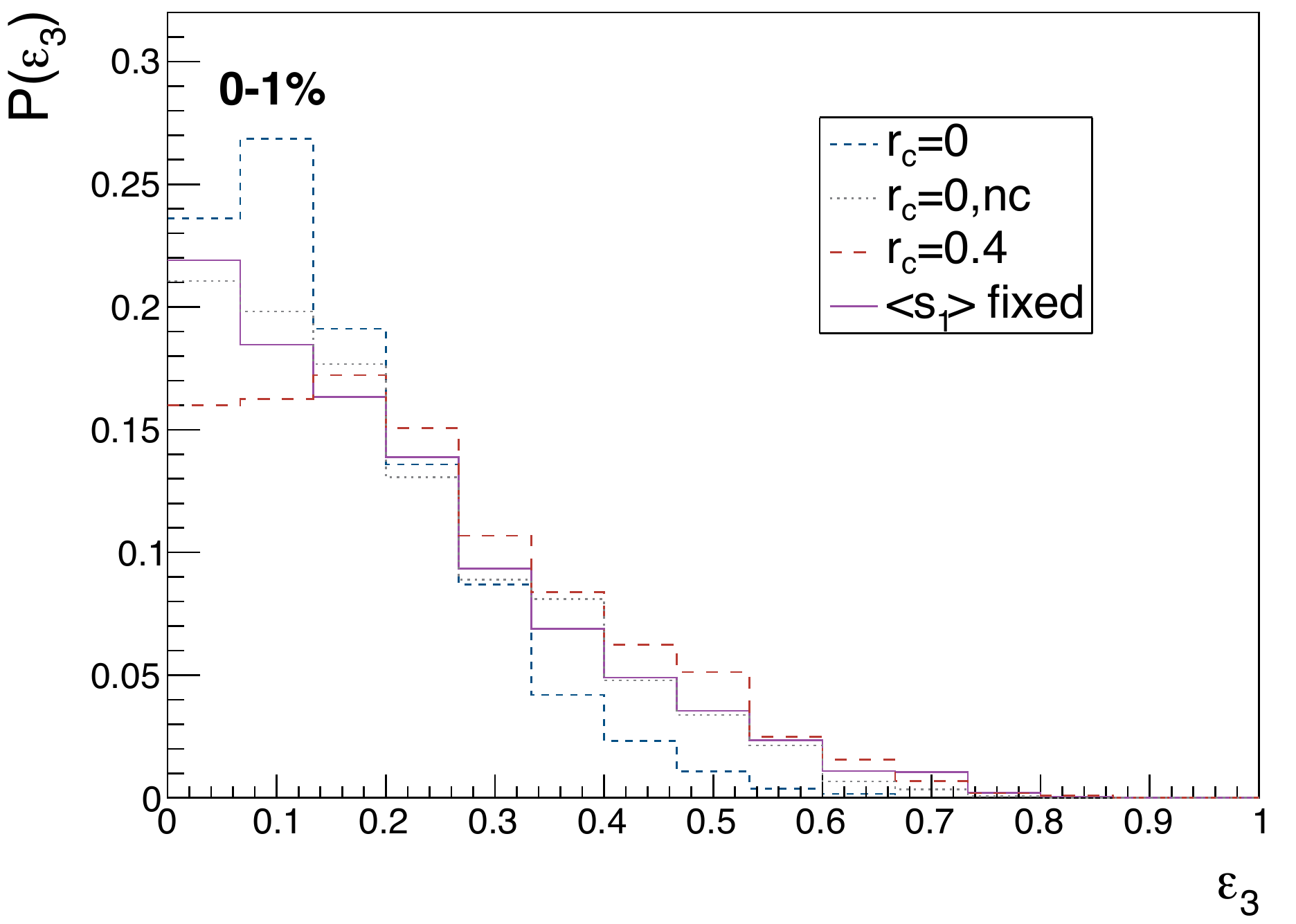} 
\end{center}
\vspace*{-0.7cm}
\caption[a]{Probability distribution of the triangularity, $\varepsilon_3$, for $r_c\!=\!0$ (blue short-dashed line), $r_c\!=\!0.4$~fm (red long-dashed line), $r_c\!=\!0,nc$ (grey dotted line) and  $\langle s_1 \rangle$ fixed (purple solid line) after selecting the 1\% most \textit{entropic} events.}
\label{eps3_01}
\end{figure}

In Figs.~\ref{eps2_01}-\ref{eps3_01} we present the calculation of the probability distributions $\mathcal{P}(\varepsilon_{2,3})$ after imposing a cut on the entropy deposition $S$ i.e. only considering the events of the $0\!-\!1$\% centrality class as given by Fig.~\ref{entropy_deposition}. Two remarkable results can be extracted by comparing Figs.~\ref{eps2}-\ref{eps3} (minimum bias) with Figs.~\ref{eps2_01}-\ref{eps3_01} (ultra-central collisions). First, we observe how the probability distributions $\mathcal{P}(\varepsilon_{2,3})$ are shifted towards larger values when selecting ultra-central events both in the uncorrelated and correlated scenarios. Next, focusing on the role of correlated constituents in this high-entropy context we observe that it turns out to favor higher values of $\varepsilon_2$ and $\varepsilon_3$ when compared to the uncorrelated scenario. Thus, we find that the consequence of having correlated constituents inside the proton is the opposite in ultra-central collisions than in minimum bias. To sum up, in the $0\!-\!1$\% centrality class the net effect of correlations is to increase the probability of having larger values of $\varepsilon_{2,3}$ whereas in the minimum bias case this probability is diminished.

\begin{figure}[h]
\begin{center}
\includegraphics[scale=0.465]{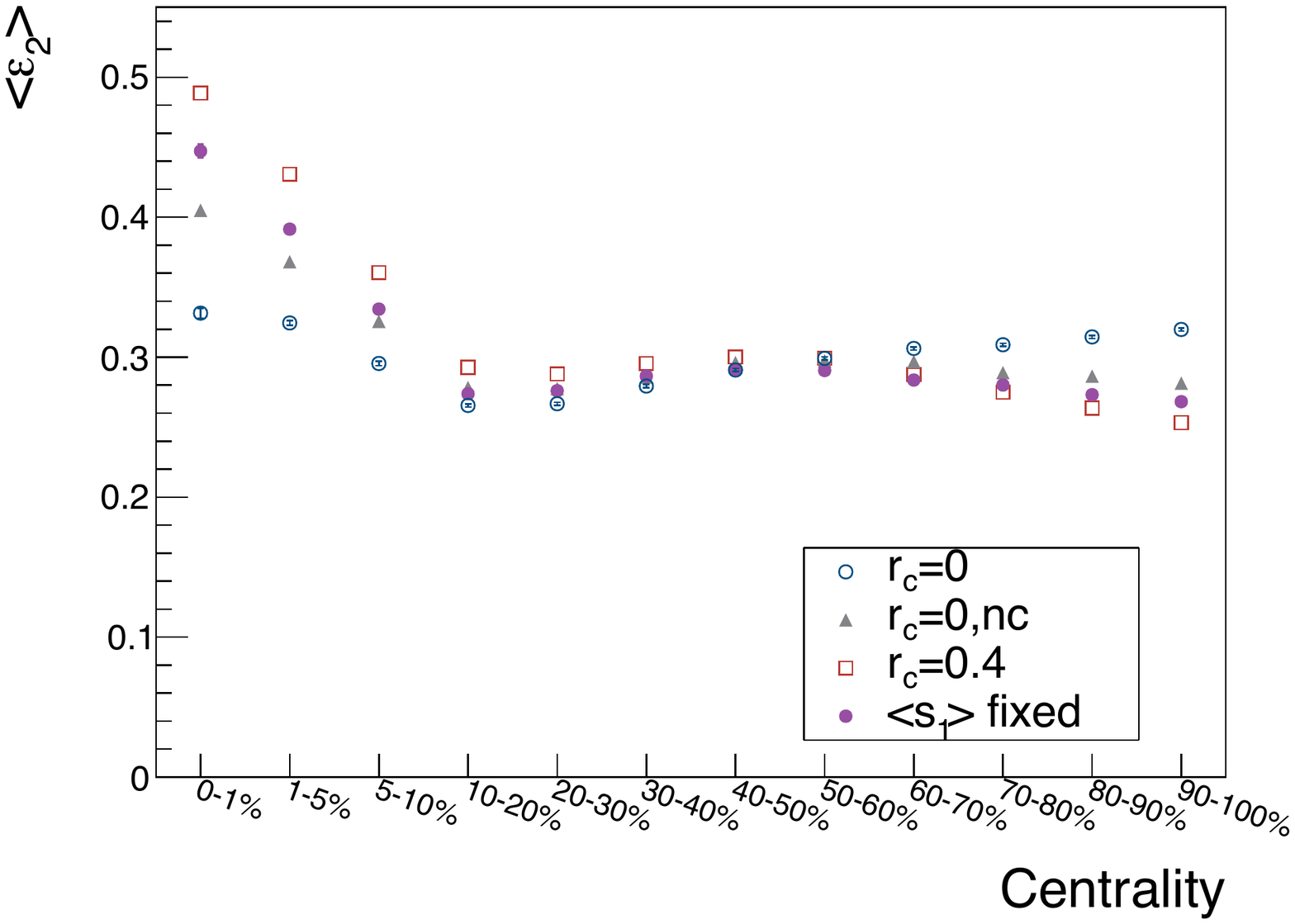} 
\end{center}
\vspace*{-0.7cm}
\caption[a]{Average values of the eccentricity, $\varepsilon_2$, for $r_c\!=\!0$ (blue empty circle), $r_c\!=\!0.4$~fm (red empty square), $r_c\!=\!0,nc$ (grey filled triangle) and  $\langle s_1 \rangle$ fixed (purple filled circle) as a function of the centrality range.}
\label{eps2_centrality}
\end{figure}

\begin{figure}[h]
\begin{center}
\includegraphics[scale=0.465]{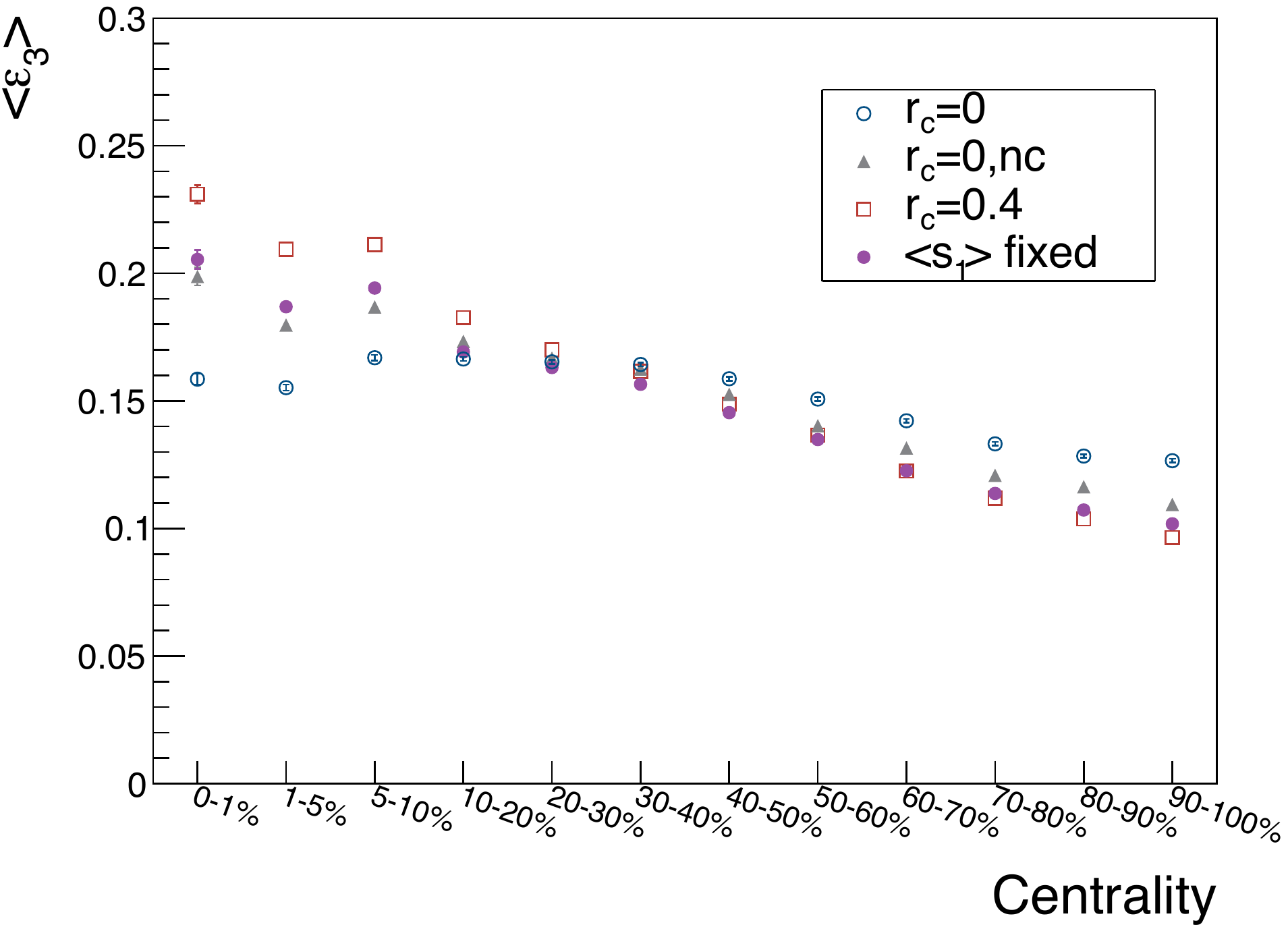} 
\end{center}
\vspace*{-0.7cm}
\caption[a]{Average values of the triangularity, $\varepsilon_3$, for $r_c\!=\!0$ (blue empty circle), $r_c\!=\!0.4$~fm (red empty square), $r_c\!=\!0,nc$ (grey filled triangle) and  $\langle s_1 \rangle$ fixed (purple filled circle) as a function of the centrality range.}
\label{eps3_centrality}
\end{figure}

The difference on the effect of spatial correlations between the minimum bias case and the ultra-central can be neatly deduced from Figs.\ref{eps2_centrality}-\ref{eps3_centrality}. We represent the average values of $\varepsilon_{2,3}$ for different centrality classes. A common trend is observed both in the correlated and uncorrelated cases: while $\langle\varepsilon_{2,3}\rangle$ is barely centrality dependent in the mid-central to peripheral collisions it increases significantly in the very central region i.e. in the events with higher entropy deposition.
Regarding the effect of spatial correlations we notice that they increase $\langle\varepsilon_{2,3}\rangle$ for the higher entropic events with respect to the uncorrelated cases as we already shown in Figs.~\ref{eps2_01}-\ref{eps3_01}. Minimum bias collisions are not dominated by these infrequent extremely entropic events but by the ones with a smaller entropy production. In this case, we see in
Figs.\ref{eps2_centrality}-\ref{eps3_centrality} how the net effect of correlations in the peripheral/less entropic bins is to reduce $\langle\varepsilon_{2,3}\rangle$. Furthermore, the quantitative difference in $\langle\varepsilon_{2,3}\rangle$ between the correlated and uncorrelated scenarios is larger in the ultra-central events than in minimum bias. Thus, we conclude that the net effect of correlated constituents is larger in ultra-central collisions.

\begin{figure}[h]
\begin{center}
\includegraphics[scale=0.465]{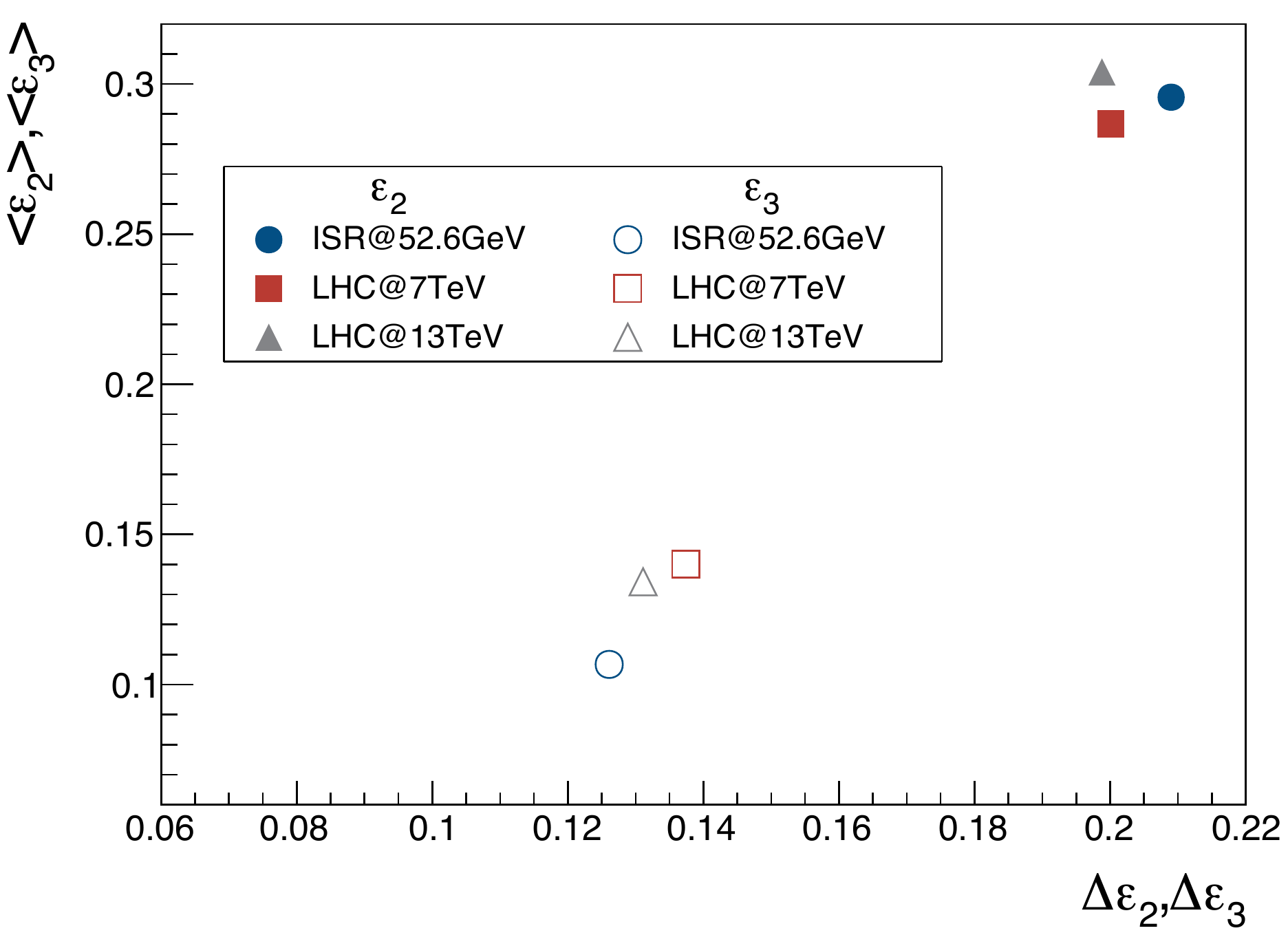} 
\end{center}
\vspace*{-0.5cm}
\caption[a]{Average values of $\varepsilon_2$ (filled markers) and $\varepsilon_3$ (empty markers) for $\sqrt s\!=\!52.6$~GeV (blue filled/empty circles), $7$~TeV (red filled/empty squares) and $13$~TeV (grey filled/empty triangles) as a function of their standard deviation. These energies correspond to ISR and Runs I-II of the LHC, respectively.}
\label{eps23_energy}
\end{figure}

\subsection{Energy scan in minimum bias}\label{es}
Before presenting our results for the energy dependence of the spatial eccentricities, it is worth to mention that, in agreement with \cite{Bozek:2016kpf}, we find moderate variations in the mean number of wounded hot spots as the energy increases. Indeed, for the $r_c\!=\!0.4$ case we obtain $\overline{N}_w\!=\!2.3$, $2.74$, $2.75$ at $\sqrt s\!=\!52.6$~GeV, $7$~TeV and $13$~TeV respectively. The rising behavior of $\overline{N}_w$ with increasing collision energy can be directly attributed to the growth of $R_{hs}$ as depicted in Table~\ref{param}. In other words, in a geometrical framework as it is the Monte-Carlo Glauber model, bigger hot spots translate into more collisions between them.

In Fig.~\ref{eps23_energy} we represent the average values of $\varepsilon_2$ and $\varepsilon_3$ as a function of their standard deviation for 3 different collision energies namely $\sqrt s\!=\!52.6,7000,13000$~GeV. All the curves refer to the $r_c\!=\!0.4$ case and taking into account all the events.
It should be noted that the energy dependence of the parameters of the model (see Table~\ref{param}) comes from the requirement of reproducing the total $pp$ cross section, being this a quite soft condition. Endowing our model with a more rigorous and precise energy dependence is left for future work. The main goal of Fig.~\ref{eps23_energy} is to show that there are no significant differences in the values of $\langle\varepsilon_2\rangle$ and $\langle\varepsilon_3\rangle$ for different collision energies. The fact that the spatial eccentricities do not drastically deviate with increasing energy was also observed in \cite{Petersen:2011sb} where a different parametrization of the initial state was used. Further, the purpose of representing the average values as a function of the standard deviation is to emphasize the width of the probability distributions that we obtain as we have seen in the previous sections (see Figs.~\ref{eps2}-\ref{eps3_01}).  

\section{Conclusions}
To conclude, in this work we present a quantitative analysis of the effect of non-trivial spatial correlations between constituents of the proton on the features of the initial state geometry in proton-proton collisions. The inclusion of correlations between subnucleonic degrees of freedom is motivated by their essential role in a plausible dynamical explanation of a new feature of $pp$ scattering observed at $\sqrt s\!=\!7$~TeV at the LHC, namely the hollowness effect. Our approach is based on a Monte-Carlo Glauber calculation that allow us to compute observables on an even-by-event basis. We follow similar steps to the ones in previous works without the correlations aforementioned. We focus on the role of the repulsive correlations and left the improvement of the physical description of processes such as the coherent particle production for future work. Essentially, the presence of correlations affect the geometry of the collision leading to variations in the basic elements of the Monte-Carlo Glauber model such as the mean number of wounded hot spots or their radial distribution in the transverse plane. In particular, we show that these correlations may produce a notorious reduction both of the eccentricity and on the triangularity in minimum bias events. However, when a cut in the entropy deposition is applied to select the 1\% of most entropic events the effect of correlations is the opposite: larger eccentricities and triangularities are expected in the correlated scenario than in the uncorrelated one in ultra-central collisions. We attribute this result to the fact that in our model the entropy deposition is computed as an incoherent superposition (see Eq.~\ref{entropy_eq}) of the individual contributions of each wounded hot spot, the more wounded hot spots the larger the entropy deposited. Then, upon imposing this high-entropy cut we are implicitly selecting events with a large number of wounded hot spots. In these configurations the spatial correlations increase the values of $\varepsilon_2$ and $\varepsilon_3$ with respect to the uncorrelated cases. In addition to the effect of correlations on the properties of the initial state we have explored their energy dependence from ISR to the LHC finding small deviations in the values of $\varepsilon_2$ and $\varepsilon_3$. The results presented in this article shall be regarded as the first step for future studies in which we would like to address the hydrodynamic evolution of the obtained initial entropy densities and the extension of the model to bigger collision systems such as $pA$ and $AA$.

\section*{ACKNOWLEDGMENTS} AS would like to thank Harri Niemi, Patricia S\'anchez-Lucas and Jos\'e I. Illana for fruitful discussions. This work was partially supported by a Helmholtz Young Investigator Group VH-NG-822 from the Helmholtz Association and GSI, the Helmholtz International
Center for the Facility for Antiproton and Ion Research (HIC for FAIR) within the
framework of the Landes-Offensive zur Entwicklung Wissenschaftlich-Oekonomischer Exzellenz (LOEWE) program launched by the State of Hesse, a FP7-PEOPLE-2013-CIG Grant of the European Commission, reference QCDense/631558, and by Ram\'on y Cajal and MINECO projects reference RYC-2011-09010 and FPA2013-47836.

\bibliography{refB}{}

\begin{thebibliography}{45}%
\makeatletter
\providecommand \@ifxundefined [1]{%
 \@ifx{#1\undefined}
}%
\providecommand \@ifnum [1]{%
 \ifnum #1\expandafter \@firstoftwo
 \else \expandafter \@secondoftwo
 \fi
}%
\providecommand \@ifx [1]{%
 \ifx #1\expandafter \@firstoftwo
 \else \expandafter \@secondoftwo
 \fi
}%
\providecommand \natexlab [1]{#1}%
\providecommand \enquote  [1]{``#1''}%
\providecommand \bibnamefont  [1]{#1}%
\providecommand \bibfnamefont [1]{#1}%
\providecommand \citenamefont [1]{#1}%
\providecommand \href@noop [0]{\@secondoftwo}%
\providecommand \href [0]{\begingroup \@sanitize@url \@href}%
\providecommand \@href[1]{\@@startlink{#1}\@@href}%
\providecommand \@@href[1]{\endgroup#1\@@endlink}%
\providecommand \@sanitize@url [0]{\catcode `\\12\catcode `\$12\catcode
  `\&12\catcode `\#12\catcode `\^12\catcode `\_12\catcode `\%12\relax}%
\providecommand \@@startlink[1]{}%
\providecommand \@@endlink[0]{}%
\providecommand \url  [0]{\begingroup\@sanitize@url \@url }%
\providecommand \@url [1]{\endgroup\@href {#1}{\urlprefix }}%
\providecommand \urlprefix  [0]{URL }%
\providecommand \Eprint [0]{\href }%
\providecommand \doibase [0]{http://dx.doi.org/}%
\providecommand \selectlanguage [0]{\@gobble}%
\providecommand \bibinfo  [0]{\@secondoftwo}%
\providecommand \bibfield  [0]{\@secondoftwo}%
\providecommand \translation [1]{[#1]}%
\providecommand \BibitemOpen [0]{}%
\providecommand \bibitemStop [0]{}%
\providecommand \bibitemNoStop [0]{.\EOS\space}%
\providecommand \EOS [0]{\spacefactor3000\relax}%
\providecommand \BibitemShut  [1]{\csname bibitem#1\endcsname}%
\let\auto@bib@innerbib\@empty
\bibitem [{\citenamefont {Arsene}\ \emph {et~al.}(2005)\citenamefont {Arsene}
  \emph {et~al.}}]{Arsene:2004fa}%
  \BibitemOpen
  \bibfield  {author} {\bibinfo {author} {\bibfnamefont {I.}~\bibnamefont
  {Arsene}} \emph {et~al.} (\bibinfo {collaboration} {BRAHMS}),\ }\href
  {\doibase 10.1016/j.nuclphysa.2005.02.130} {\bibfield  {journal} {\bibinfo
  {journal} {Nucl. Phys.}\ }\textbf {\bibinfo {volume} {A757}},\ \bibinfo
  {pages} {1} (\bibinfo {year} {2005})}\BibitemShut {NoStop}%
\bibitem [{\citenamefont {Aamodt}\ \emph
  {et~al.}(2010{\natexlab{a}})\citenamefont {Aamodt} \emph
  {et~al.}}]{Aamodt:2010pa}%
  \BibitemOpen
  \bibfield  {author} {\bibinfo {author} {\bibfnamefont {K.}~\bibnamefont
  {Aamodt}} \emph {et~al.} (\bibinfo {collaboration} {ALICE}),\ }\href
  {\doibase 10.1103/PhysRevLett.105.252302} {\bibfield  {journal} {\bibinfo
  {journal} {Phys. Rev. Lett.}\ }\textbf {\bibinfo {volume} {105}},\ \bibinfo
  {pages} {252302} (\bibinfo {year} {2010}{\natexlab{a}})}\BibitemShut
  {NoStop}%
\bibitem [{\citenamefont {Adam}\ \emph {et~al.}(2016)\citenamefont {Adam} \emph
  {et~al.}}]{Adam:2016izf}%
  \BibitemOpen
  \bibfield  {author} {\bibinfo {author} {\bibfnamefont {J.}~\bibnamefont
  {Adam}} \emph {et~al.} (\bibinfo {collaboration} {ALICE}),\ }\href {\doibase
  10.1103/PhysRevLett.116.132302} {\bibfield  {journal} {\bibinfo  {journal}
  {Phys. Rev. Lett.}\ }\textbf {\bibinfo {volume} {116}},\ \bibinfo {pages}
  {132302} (\bibinfo {year} {2016})}\BibitemShut {NoStop}%
\bibitem [{\citenamefont {Khachatryan}\ \emph {et~al.}(2010)\citenamefont
  {Khachatryan} \emph {et~al.}}]{Khachatryan:2010gv}%
  \BibitemOpen
  \bibfield  {author} {\bibinfo {author} {\bibfnamefont {V.}~\bibnamefont
  {Khachatryan}} \emph {et~al.} (\bibinfo {collaboration} {CMS}),\ }\href
  {\doibase 10.1007/JHEP09(2010)091} {\bibfield  {journal} {\bibinfo  {journal}
  {JHEP}\ }\textbf {\bibinfo {volume} {09}},\ \bibinfo {pages} {091} (\bibinfo
  {year} {2010})}\BibitemShut {NoStop}%
\bibitem [{\citenamefont {Aad}\ \emph {et~al.}(2016)\citenamefont {Aad} \emph
  {et~al.}}]{Aad:2015gqa}%
  \BibitemOpen
  \bibfield  {author} {\bibinfo {author} {\bibfnamefont {G.}~\bibnamefont
  {Aad}} \emph {et~al.} (\bibinfo {collaboration} {ATLAS}),\ }\href {\doibase
  10.1103/PhysRevLett.116.172301} {\bibfield  {journal} {\bibinfo  {journal}
  {Phys. Rev. Lett.}\ }\textbf {\bibinfo {volume} {116}},\ \bibinfo {pages}
  {172301} (\bibinfo {year} {2016})}\BibitemShut {NoStop}%
\bibitem [{\citenamefont {Khachatryan}\ \emph {et~al.}(2017)\citenamefont
  {Khachatryan} \emph {et~al.}}]{Khachatryan:2016txc}%
  \BibitemOpen
  \bibfield  {author} {\bibinfo {author} {\bibfnamefont {V.}~\bibnamefont
  {Khachatryan}} \emph {et~al.} (\bibinfo {collaboration} {CMS}),\ }\href
  {\doibase 10.1016/j.physletb.2016.12.009} {\bibfield  {journal} {\bibinfo
  {journal} {Phys. Lett.}\ }\textbf {\bibinfo {volume} {B765}},\ \bibinfo
  {pages} {193} (\bibinfo {year} {2017})}\BibitemShut {NoStop}%
\bibitem [{\citenamefont {Weller}\ and\ \citenamefont
  {Romatschke}(2017)}]{Weller:2017tsr}%
  \BibitemOpen
  \bibfield  {author} {\bibinfo {author} {\bibfnamefont {R.~D.}\ \bibnamefont
  {Weller}}\ and\ \bibinfo {author} {\bibfnamefont {P.}~\bibnamefont
  {Romatschke}},\ }\href@noop {} {\  (\bibinfo {year} {2017})},\ \Eprint
  {http://arxiv.org/abs/1701.07145} {arXiv:1701.07145 [nucl-th]} \BibitemShut
  {NoStop}%
\bibitem [{\citenamefont {Schlichting}\ and\ \citenamefont
  {Tribedy}(2016)}]{Schlichting:2016kjw}%
  \BibitemOpen
  \bibfield  {author} {\bibinfo {author} {\bibfnamefont {S.}~\bibnamefont
  {Schlichting}}\ and\ \bibinfo {author} {\bibfnamefont {P.}~\bibnamefont
  {Tribedy}},\ }\href {\doibase 10.1155/2016/8460349} {\bibfield  {journal}
  {\bibinfo  {journal} {Adv. High Energy Phys.}\ }\textbf {\bibinfo {volume}
  {2016}},\ \bibinfo {pages} {8460349} (\bibinfo {year} {2016})}\BibitemShut
  {NoStop}%
\bibitem [{\citenamefont {M{\"a}ntysaari}\ and\ \citenamefont
  {Schenke}(2016)}]{Mantysaari:2016ykx}%
  \BibitemOpen
  \bibfield  {author} {\bibinfo {author} {\bibfnamefont {H.}~\bibnamefont
  {M{\"a}ntysaari}}\ and\ \bibinfo {author} {\bibfnamefont {B.}~\bibnamefont
  {Schenke}},\ }\href {\doibase 10.1103/PhysRevLett.117.052301} {\bibfield
  {journal} {\bibinfo  {journal} {Phys. Rev. Lett.}\ }\textbf {\bibinfo
  {volume} {117}},\ \bibinfo {pages} {052301} (\bibinfo {year}
  {2016})}\BibitemShut {NoStop}%
\bibitem [{\citenamefont {Schlichting}\ and\ \citenamefont
  {Schenke}(2014)}]{Schlichting:2014ipa}%
  \BibitemOpen
  \bibfield  {author} {\bibinfo {author} {\bibfnamefont {S.}~\bibnamefont
  {Schlichting}}\ and\ \bibinfo {author} {\bibfnamefont {B.}~\bibnamefont
  {Schenke}},\ }\href {\doibase 10.1016/j.physletb.2014.10.068} {\bibfield
  {journal} {\bibinfo  {journal} {Phys. Lett.}\ }\textbf {\bibinfo {volume}
  {B739}},\ \bibinfo {pages} {313} (\bibinfo {year} {2014})}\BibitemShut
  {NoStop}%
\bibitem [{\citenamefont {Loizides}(2016)}]{Loizides:2016djv}%
  \BibitemOpen
  \bibfield  {author} {\bibinfo {author} {\bibfnamefont {C.}~\bibnamefont
  {Loizides}},\ }\href {\doibase 10.1103/PhysRevC.94.024914} {\bibfield
  {journal} {\bibinfo  {journal} {Phys. Rev.}\ }\textbf {\bibinfo {volume}
  {C94}},\ \bibinfo {pages} {024914} (\bibinfo {year} {2016})}\BibitemShut
  {NoStop}%
\bibitem [{\citenamefont {Bo{\.z}ek}\ \emph {et~al.}(2016)\citenamefont
  {Bo{\.z}ek}, \citenamefont {Broniowski},\ and\ \citenamefont
  {Rybczy{\'n}ski}}]{Bozek:2016kpf}%
  \BibitemOpen
  \bibfield  {author} {\bibinfo {author} {\bibfnamefont {P.}~\bibnamefont
  {Bo{\.z}ek}}, \bibinfo {author} {\bibfnamefont {W.}~\bibnamefont
  {Broniowski}}, \ and\ \bibinfo {author} {\bibfnamefont {M.}~\bibnamefont
  {Rybczy{\'n}ski}},\ }\href {\doibase 10.1103/PhysRevC.94.014902} {\bibfield
  {journal} {\bibinfo  {journal} {Phys. Rev.}\ }\textbf {\bibinfo {volume}
  {C94}},\ \bibinfo {pages} {014902} (\bibinfo {year} {2016})}\BibitemShut
  {NoStop}%
\bibitem [{\citenamefont {Welsh}\ \emph {et~al.}(2016)\citenamefont {Welsh},
  \citenamefont {Singer},\ and\ \citenamefont {Heinz}}]{Welsh:2016siu}%
  \BibitemOpen
  \bibfield  {author} {\bibinfo {author} {\bibfnamefont {K.}~\bibnamefont
  {Welsh}}, \bibinfo {author} {\bibfnamefont {J.}~\bibnamefont {Singer}}, \
  and\ \bibinfo {author} {\bibfnamefont {U.~W.}\ \bibnamefont {Heinz}},\ }\href
  {\doibase 10.1103/PhysRevC.94.024919} {\bibfield  {journal} {\bibinfo
  {journal} {Phys. Rev.}\ }\textbf {\bibinfo {volume} {C94}},\ \bibinfo {pages}
  {024919} (\bibinfo {year} {2016})}\BibitemShut {NoStop}%
\bibitem [{\citenamefont {Dumitru}\ and\ \citenamefont
  {Nara}(2012)}]{Dumitru:2012yr}%
  \BibitemOpen
  \bibfield  {author} {\bibinfo {author} {\bibfnamefont {A.}~\bibnamefont
  {Dumitru}}\ and\ \bibinfo {author} {\bibfnamefont {Y.}~\bibnamefont {Nara}},\
  }\href {\doibase 10.1103/PhysRevC.85.034907} {\bibfield  {journal} {\bibinfo
  {journal} {Phys. Rev.}\ }\textbf {\bibinfo {volume} {C85}},\ \bibinfo {pages}
  {034907} (\bibinfo {year} {2012})}\BibitemShut {NoStop}%
\bibitem [{\citenamefont {Miller}\ and\ \citenamefont
  {Snellings}(2003)}]{Miller:2003kd}%
  \BibitemOpen
  \bibfield  {author} {\bibinfo {author} {\bibfnamefont {M.}~\bibnamefont
  {Miller}}\ and\ \bibinfo {author} {\bibfnamefont {R.}~\bibnamefont
  {Snellings}},\ }\href@noop {} {\  (\bibinfo {year} {2003})},\ \Eprint
  {http://arxiv.org/abs/nucl-ex/0312008} {arXiv:nucl-ex/0312008 [nucl-ex]}
  \BibitemShut {NoStop}%
\bibitem [{\citenamefont {Schenke}\ \emph
  {et~al.}(2012{\natexlab{a}})\citenamefont {Schenke}, \citenamefont
  {Tribedy},\ and\ \citenamefont {Venugopalan}}]{Schenke:2012fw}%
  \BibitemOpen
  \bibfield  {author} {\bibinfo {author} {\bibfnamefont {B.}~\bibnamefont
  {Schenke}}, \bibinfo {author} {\bibfnamefont {P.}~\bibnamefont {Tribedy}}, \
  and\ \bibinfo {author} {\bibfnamefont {R.}~\bibnamefont {Venugopalan}},\
  }\href {\doibase 10.1103/PhysRevC.86.034908} {\bibfield  {journal} {\bibinfo
  {journal} {Phys. Rev.}\ }\textbf {\bibinfo {volume} {C86}},\ \bibinfo {pages}
  {034908} (\bibinfo {year} {2012}{\natexlab{a}})}\BibitemShut {NoStop}%
\bibitem [{\citenamefont {Gr{\"o}nqvist}\ \emph {et~al.}(2016)\citenamefont
  {Gr{\"o}nqvist}, \citenamefont {Blaizot},\ and\ \citenamefont
  {Ollitrault}}]{Gronqvist:2016hym}%
  \BibitemOpen
  \bibfield  {author} {\bibinfo {author} {\bibfnamefont {H.}~\bibnamefont
  {Gr{\"o}nqvist}}, \bibinfo {author} {\bibfnamefont {J.-P.}\ \bibnamefont
  {Blaizot}}, \ and\ \bibinfo {author} {\bibfnamefont {J.-Y.}\ \bibnamefont
  {Ollitrault}},\ }\href {\doibase 10.1103/PhysRevC.94.034905} {\bibfield
  {journal} {\bibinfo  {journal} {Phys. Rev.}\ }\textbf {\bibinfo {volume}
  {C94}},\ \bibinfo {pages} {034905} (\bibinfo {year} {2016})},\ \Eprint
  {http://arxiv.org/abs/1604.07230} {arXiv:1604.07230} \BibitemShut {NoStop}%
\bibitem [{\citenamefont {Avsar}\ \emph {et~al.}(2011)\citenamefont {Avsar},
  \citenamefont {Flensburg}, \citenamefont {Hatta}, \citenamefont
  {Ollitrault},\ and\ \citenamefont {Ueda}}]{Avsar:2010rf}%
  \BibitemOpen
  \bibfield  {author} {\bibinfo {author} {\bibfnamefont {E.}~\bibnamefont
  {Avsar}}, \bibinfo {author} {\bibfnamefont {C.}~\bibnamefont {Flensburg}},
  \bibinfo {author} {\bibfnamefont {Y.}~\bibnamefont {Hatta}}, \bibinfo
  {author} {\bibfnamefont {J.-Y.}\ \bibnamefont {Ollitrault}}, \ and\ \bibinfo
  {author} {\bibfnamefont {T.}~\bibnamefont {Ueda}},\ }\href {\doibase
  10.1016/j.physletb.2011.07.031} {\bibfield  {journal} {\bibinfo  {journal}
  {Phys. Lett.}\ }\textbf {\bibinfo {volume} {B702}},\ \bibinfo {pages} {394}
  (\bibinfo {year} {2011})}\BibitemShut {NoStop}%
\bibitem [{\citenamefont {Mäntysaari}\ \emph {et~al.}(2017)\citenamefont
  {Mäntysaari}, \citenamefont {Schenke}, \citenamefont {Shen},\ and\
  \citenamefont {Tribedy}}]{Mantysaari:2017cni}%
  \BibitemOpen
  \bibfield  {author} {\bibinfo {author} {\bibfnamefont {H.}~\bibnamefont
  {Mäntysaari}}, \bibinfo {author} {\bibfnamefont {B.}~\bibnamefont
  {Schenke}}, \bibinfo {author} {\bibfnamefont {C.}~\bibnamefont {Shen}}, \
  and\ \bibinfo {author} {\bibfnamefont {P.}~\bibnamefont {Tribedy}},\
  }\href@noop {} {\  (\bibinfo {year} {2017})},\ \Eprint
  {http://arxiv.org/abs/1705.03177} {arXiv:1705.03177 [nucl-th]} \BibitemShut
  {NoStop}%
\bibitem [{\citenamefont {Antchev}\ \emph {et~al.}(2011)\citenamefont {Antchev}
  \emph {et~al.}}]{Antchev:2011zz}%
  \BibitemOpen
  \bibfield  {author} {\bibinfo {author} {\bibfnamefont {G.}~\bibnamefont
  {Antchev}} \emph {et~al.} (\bibinfo {collaboration} {TOTEM}),\ }\href
  {\doibase 10.1209/0295-5075/95/41001} {\bibfield  {journal} {\bibinfo
  {journal} {Europhys. Lett.}\ }\textbf {\bibinfo {volume} {95}},\ \bibinfo
  {pages} {41001} (\bibinfo {year} {2011})}\BibitemShut {NoStop}%
\bibitem [{\citenamefont {Dremin}(2017)}]{Dremin:2015ujt}%
  \BibitemOpen
  \bibfield  {author} {\bibinfo {author} {\bibfnamefont {I.~M.}\ \bibnamefont
  {Dremin}},\ }\href {\doibase 10.3103/S1068335617040029} {\bibfield  {journal}
  {\bibinfo  {journal} {Bull. Lebedev Phys. Inst.}\ }\textbf {\bibinfo {volume}
  {44}},\ \bibinfo {pages} {94} (\bibinfo {year} {2017})}\BibitemShut {NoStop}%
\bibitem [{\citenamefont {Alkin}\ \emph {et~al.}(2014)\citenamefont {Alkin},
  \citenamefont {Martynov}, \citenamefont {Kovalenko},\ and\ \citenamefont
  {Troshin}}]{Alkin:2014rfa}%
  \BibitemOpen
  \bibfield  {author} {\bibinfo {author} {\bibfnamefont {A.}~\bibnamefont
  {Alkin}}, \bibinfo {author} {\bibfnamefont {E.}~\bibnamefont {Martynov}},
  \bibinfo {author} {\bibfnamefont {O.}~\bibnamefont {Kovalenko}}, \ and\
  \bibinfo {author} {\bibfnamefont {S.~M.}\ \bibnamefont {Troshin}},\ }\href
  {\doibase 10.1103/PhysRevD.89.091501} {\bibfield  {journal} {\bibinfo
  {journal} {Phys. Rev.}\ }\textbf {\bibinfo {volume} {D89}},\ \bibinfo {pages}
  {091501} (\bibinfo {year} {2014})}\BibitemShut {NoStop}%
\bibitem [{\citenamefont {Troshin}\ and\ \citenamefont
  {Tyurin}(2016)}]{Troshin:2016frs}%
  \BibitemOpen
  \bibfield  {author} {\bibinfo {author} {\bibfnamefont {S.~M.}\ \bibnamefont
  {Troshin}}\ and\ \bibinfo {author} {\bibfnamefont {N.~E.}\ \bibnamefont
  {Tyurin}},\ }\href {\doibase 10.1142/S0217732316500796} {\bibfield  {journal}
  {\bibinfo  {journal} {Mod. Phys. Lett.}\ }\textbf {\bibinfo {volume} {A31}},\
  \bibinfo {pages} {1650079} (\bibinfo {year} {2016})}\BibitemShut {NoStop}%
\bibitem [{\citenamefont {Dremin}(2016)}]{Dremin:2016ugi}%
  \BibitemOpen
  \bibfield  {author} {\bibinfo {author} {\bibfnamefont {I.~M.}\ \bibnamefont
  {Dremin}},\ }\href@noop {} {\  (\bibinfo {year} {2016})},\ \Eprint
  {http://arxiv.org/abs/1610.07937} {arXiv:1610.07937 [hep-ph]} \BibitemShut
  {NoStop}%
\bibitem [{\citenamefont {Ruiz~Arriola}\ and\ \citenamefont
  {Broniowski}(2016)}]{Arriola:2016bxa}%
  \BibitemOpen
  \bibfield  {author} {\bibinfo {author} {\bibfnamefont {E.}~\bibnamefont
  {Ruiz~Arriola}}\ and\ \bibinfo {author} {\bibfnamefont {W.}~\bibnamefont
  {Broniowski}},\ }\bibfield  {booktitle} {\emph {\bibinfo {booktitle}
  {{Proceedings, Theory and Experiment for Hadrons on the Light-Front (Light
  Cone 2015): Frascati , Italy, September 21-25, 2015}}},\ }\href {\doibase
  10.1007/s00601-016-1095-z} {\bibfield  {journal} {\bibinfo  {journal} {Few
  Body Syst.}\ }\textbf {\bibinfo {volume} {57}},\ \bibinfo {pages} {485}
  (\bibinfo {year} {2016})}\BibitemShut {NoStop}%
\bibitem [{\citenamefont {Giacalone}\ \emph {et~al.}(2017)\citenamefont
  {Giacalone}, \citenamefont {Noronha-Hostler},\ and\ \citenamefont
  {Ollitrault}}]{Giacalone:2017uqx}%
  \BibitemOpen
  \bibfield  {author} {\bibinfo {author} {\bibfnamefont {G.}~\bibnamefont
  {Giacalone}}, \bibinfo {author} {\bibfnamefont {J.}~\bibnamefont
  {Noronha-Hostler}}, \ and\ \bibinfo {author} {\bibfnamefont {J.-Y.}\
  \bibnamefont {Ollitrault}},\ }\href {\doibase 10.1103/PhysRevC.95.054910}
  {\bibfield  {journal} {\bibinfo  {journal} {Phys. Rev.}\ }\textbf {\bibinfo
  {volume} {C95}},\ \bibinfo {pages} {054910} (\bibinfo {year}
  {2017})}\BibitemShut {NoStop}%
\bibitem [{\citenamefont {Alvioli}\ \emph {et~al.}(2009)\citenamefont
  {Alvioli}, \citenamefont {Drescher},\ and\ \citenamefont
  {Strikman}}]{Alvioli:2009ab}%
  \BibitemOpen
  \bibfield  {author} {\bibinfo {author} {\bibfnamefont {M.}~\bibnamefont
  {Alvioli}}, \bibinfo {author} {\bibfnamefont {H.~J.}\ \bibnamefont
  {Drescher}}, \ and\ \bibinfo {author} {\bibfnamefont {M.}~\bibnamefont
  {Strikman}},\ }\href {\doibase 10.1016/j.physletb.2009.08.067} {\bibfield
  {journal} {\bibinfo  {journal} {Phys. Lett.}\ }\textbf {\bibinfo {volume}
  {B680}},\ \bibinfo {pages} {225} (\bibinfo {year} {2009})}\BibitemShut
  {NoStop}%
\bibitem [{\citenamefont {Blaizot}\ \emph {et~al.}(2014)\citenamefont
  {Blaizot}, \citenamefont {Broniowski},\ and\ \citenamefont
  {Ollitrault}}]{Blaizot:2014wba}%
  \BibitemOpen
  \bibfield  {author} {\bibinfo {author} {\bibfnamefont {J.-P.}\ \bibnamefont
  {Blaizot}}, \bibinfo {author} {\bibfnamefont {W.}~\bibnamefont {Broniowski}},
  \ and\ \bibinfo {author} {\bibfnamefont {J.-Y.}\ \bibnamefont {Ollitrault}},\
  }\href {\doibase 10.1103/PhysRevC.90.034906} {\bibfield  {journal} {\bibinfo
  {journal} {Phys. Rev.}\ }\textbf {\bibinfo {volume} {C90}},\ \bibinfo {pages}
  {034906} (\bibinfo {year} {2014})}\BibitemShut {NoStop}%
\bibitem [{\citenamefont {Denicol}\ \emph {et~al.}(2014)\citenamefont
  {Denicol}, \citenamefont {Gale}, \citenamefont {Jeon}, \citenamefont
  {Paquet},\ and\ \citenamefont {Schenke}}]{Denicol:2014ywa}%
  \BibitemOpen
  \bibfield  {author} {\bibinfo {author} {\bibfnamefont {G.~S.}\ \bibnamefont
  {Denicol}}, \bibinfo {author} {\bibfnamefont {C.}~\bibnamefont {Gale}},
  \bibinfo {author} {\bibfnamefont {S.}~\bibnamefont {Jeon}}, \bibinfo {author}
  {\bibfnamefont {J.~F.}\ \bibnamefont {Paquet}}, \ and\ \bibinfo {author}
  {\bibfnamefont {B.}~\bibnamefont {Schenke}},\ }\href@noop {} {\  (\bibinfo
  {year} {2014})},\ \Eprint {http://arxiv.org/abs/1406.7792} {arXiv:1406.7792
  [nucl-th]} \BibitemShut {NoStop}%
\bibitem [{\citenamefont {Broniowski}\ \emph {et~al.}(2009)\citenamefont
  {Broniowski}, \citenamefont {Rybczynski},\ and\ \citenamefont
  {Bozek}}]{Broniowski:2007nz}%
  \BibitemOpen
  \bibfield  {author} {\bibinfo {author} {\bibfnamefont {W.}~\bibnamefont
  {Broniowski}}, \bibinfo {author} {\bibfnamefont {M.}~\bibnamefont
  {Rybczynski}}, \ and\ \bibinfo {author} {\bibfnamefont {P.}~\bibnamefont
  {Bozek}},\ }\href {\doibase 10.1016/j.cpc.2008.07.016} {\bibfield  {journal}
  {\bibinfo  {journal} {Comput. Phys. Commun.}\ }\textbf {\bibinfo {volume}
  {180}},\ \bibinfo {pages} {69} (\bibinfo {year} {2009})}\BibitemShut
  {NoStop}%
\bibitem [{\citenamefont {Alver}\ \emph {et~al.}(2008)\citenamefont {Alver},
  \citenamefont {Baker}, \citenamefont {Loizides},\ and\ \citenamefont
  {Steinberg}}]{Alver:2008aq}%
  \BibitemOpen
  \bibfield  {author} {\bibinfo {author} {\bibfnamefont {B.}~\bibnamefont
  {Alver}}, \bibinfo {author} {\bibfnamefont {M.}~\bibnamefont {Baker}},
  \bibinfo {author} {\bibfnamefont {C.}~\bibnamefont {Loizides}}, \ and\
  \bibinfo {author} {\bibfnamefont {P.}~\bibnamefont {Steinberg}},\ }\href@noop
  {} {\  (\bibinfo {year} {2008})},\ \Eprint {http://arxiv.org/abs/0805.4411}
  {arXiv:0805.4411 [nucl-ex]} \BibitemShut {NoStop}%
\bibitem [{\citenamefont {Schenke}\ \emph
  {et~al.}(2012{\natexlab{b}})\citenamefont {Schenke}, \citenamefont
  {Tribedy},\ and\ \citenamefont {Venugopalan}}]{Schenke:2012wb}%
  \BibitemOpen
  \bibfield  {author} {\bibinfo {author} {\bibfnamefont {B.}~\bibnamefont
  {Schenke}}, \bibinfo {author} {\bibfnamefont {P.}~\bibnamefont {Tribedy}}, \
  and\ \bibinfo {author} {\bibfnamefont {R.}~\bibnamefont {Venugopalan}},\
  }\href {\doibase 10.1103/PhysRevLett.108.252301} {\bibfield  {journal}
  {\bibinfo  {journal} {Phys. Rev. Lett.}\ }\textbf {\bibinfo {volume} {108}},\
  \bibinfo {pages} {252301} (\bibinfo {year} {2012}{\natexlab{b}})}\BibitemShut
  {NoStop}%
\bibitem [{\citenamefont {Albacete}\ \emph {et~al.}(2011)\citenamefont
  {Albacete}, \citenamefont {Dumitru},\ and\ \citenamefont
  {Nara}}]{Albacete:2011fw}%
  \BibitemOpen
  \bibfield  {author} {\bibinfo {author} {\bibfnamefont {J.~L.}\ \bibnamefont
  {Albacete}}, \bibinfo {author} {\bibfnamefont {A.}~\bibnamefont {Dumitru}}, \
  and\ \bibinfo {author} {\bibfnamefont {Y.}~\bibnamefont {Nara}},\ }\bibfield
  {booktitle} {\emph {\bibinfo {booktitle} {{Proceedings, 27th Winter Workshop
  on Nuclear Physics (WWND 2011: Winter Park, USA, February 6-13. 2011}}},\
  }\href {\doibase 10.1088/1742-6596/316/1/012011} {\bibfield  {journal}
  {\bibinfo  {journal} {J. Phys. Conf. Ser.}\ }\textbf {\bibinfo {volume}
  {316}},\ \bibinfo {pages} {012011} (\bibinfo {year} {2011})}\BibitemShut
  {NoStop}%
\bibitem [{\citenamefont {Albacete}\ \emph {et~al.}(2013)\citenamefont
  {Albacete}, \citenamefont {Dumitru}, \citenamefont {Fujii},\ and\
  \citenamefont {Nara}}]{Albacete:2012xq}%
  \BibitemOpen
  \bibfield  {author} {\bibinfo {author} {\bibfnamefont {J.~L.}\ \bibnamefont
  {Albacete}}, \bibinfo {author} {\bibfnamefont {A.}~\bibnamefont {Dumitru}},
  \bibinfo {author} {\bibfnamefont {H.}~\bibnamefont {Fujii}}, \ and\ \bibinfo
  {author} {\bibfnamefont {Y.}~\bibnamefont {Nara}},\ }\href {\doibase
  10.1016/j.nuclphysa.2012.09.012} {\bibfield  {journal} {\bibinfo  {journal}
  {Nucl. Phys.}\ }\textbf {\bibinfo {volume} {A897}},\ \bibinfo {pages} {1}
  (\bibinfo {year} {2013})},\ \Eprint {http://arxiv.org/abs/1209.2001}
  {arXiv:1209.2001} \BibitemShut {NoStop}%
\bibitem [{\citenamefont {Bialas}\ \emph {et~al.}(1977)\citenamefont {Bialas},
  \citenamefont {Czyz},\ and\ \citenamefont {Furmanski}}]{Bialas:1977en}%
  \BibitemOpen
  \bibfield  {author} {\bibinfo {author} {\bibfnamefont {A.}~\bibnamefont
  {Bialas}}, \bibinfo {author} {\bibfnamefont {W.}~\bibnamefont {Czyz}}, \ and\
  \bibinfo {author} {\bibfnamefont {W.}~\bibnamefont {Furmanski}},\ }\href@noop
  {} {\bibfield  {journal} {\bibinfo  {journal} {Acta Phys. Polon.}\ }\textbf
  {\bibinfo {volume} {B8}},\ \bibinfo {pages} {585} (\bibinfo {year}
  {1977})}\BibitemShut {NoStop}%
\bibitem [{\citenamefont {Bialas}\ \emph {et~al.}(1976)\citenamefont {Bialas},
  \citenamefont {Bleszynski},\ and\ \citenamefont {Czyz}}]{Bialas:1976ed}%
  \BibitemOpen
  \bibfield  {author} {\bibinfo {author} {\bibfnamefont {A.}~\bibnamefont
  {Bialas}}, \bibinfo {author} {\bibfnamefont {M.}~\bibnamefont {Bleszynski}},
  \ and\ \bibinfo {author} {\bibfnamefont {W.}~\bibnamefont {Czyz}},\ }\href
  {\doibase 10.1016/0550-3213(76)90329-1} {\bibfield  {journal} {\bibinfo
  {journal} {Nucl. Phys.}\ }\textbf {\bibinfo {volume} {B111}},\ \bibinfo
  {pages} {461} (\bibinfo {year} {1976})}\BibitemShut {NoStop}%
\bibitem [{\citenamefont {Amaldi}\ and\ \citenamefont
  {Schubert}(1980)}]{Amaldi:1979kd}%
  \BibitemOpen
  \bibfield  {author} {\bibinfo {author} {\bibfnamefont {U.}~\bibnamefont
  {Amaldi}}\ and\ \bibinfo {author} {\bibfnamefont {K.~R.}\ \bibnamefont
  {Schubert}},\ }\href {\doibase 10.1016/0550-3213(80)90229-1} {\bibfield
  {journal} {\bibinfo  {journal} {Nucl. Phys.}\ }\textbf {\bibinfo {volume}
  {B166}},\ \bibinfo {pages} {301} (\bibinfo {year} {1980})}\BibitemShut
  {NoStop}%
\bibitem [{\citenamefont {Cudell}\ \emph {et~al.}(2002)\citenamefont {Cudell},
  \citenamefont {Ezhela}, \citenamefont {Gauron}, \citenamefont {Kang},
  \citenamefont {Kuyanov}, \citenamefont {Lugovsky}, \citenamefont {Martynov},
  \citenamefont {Nicolescu}, \citenamefont {Razuvaev},\ and\ \citenamefont
  {Tkachenko}}]{Cudell:2002xe}%
  \BibitemOpen
  \bibfield  {author} {\bibinfo {author} {\bibfnamefont {J.~R.}\ \bibnamefont
  {Cudell}}, \bibinfo {author} {\bibfnamefont {V.~V.}\ \bibnamefont {Ezhela}},
  \bibinfo {author} {\bibfnamefont {P.}~\bibnamefont {Gauron}}, \bibinfo
  {author} {\bibfnamefont {K.}~\bibnamefont {Kang}}, \bibinfo {author}
  {\bibfnamefont {{\relax Yu}.~V.}\ \bibnamefont {Kuyanov}}, \bibinfo {author}
  {\bibfnamefont {S.~B.}\ \bibnamefont {Lugovsky}}, \bibinfo {author}
  {\bibfnamefont {E.}~\bibnamefont {Martynov}}, \bibinfo {author}
  {\bibfnamefont {B.}~\bibnamefont {Nicolescu}}, \bibinfo {author}
  {\bibfnamefont {E.~A.}\ \bibnamefont {Razuvaev}}, \ and\ \bibinfo {author}
  {\bibfnamefont {N.~P.}\ \bibnamefont {Tkachenko}} (\bibinfo {collaboration}
  {COMPETE}),\ }\href {\doibase 10.1103/PhysRevLett.89.201801} {\bibfield
  {journal} {\bibinfo  {journal} {Phys. Rev. Lett.}\ }\textbf {\bibinfo
  {volume} {89}},\ \bibinfo {pages} {201801} (\bibinfo {year}
  {2002})}\BibitemShut {NoStop}%
\bibitem [{\citenamefont {Qiu}\ and\ \citenamefont {Heinz}(2011)}]{Qiu:2011iv}%
  \BibitemOpen
  \bibfield  {author} {\bibinfo {author} {\bibfnamefont {Z.}~\bibnamefont
  {Qiu}}\ and\ \bibinfo {author} {\bibfnamefont {U.~W.}\ \bibnamefont
  {Heinz}},\ }\href {\doibase 10.1103/PhysRevC.84.024911} {\bibfield  {journal}
  {\bibinfo  {journal} {Phys. Rev.}\ }\textbf {\bibinfo {volume} {C84}},\
  \bibinfo {pages} {024911} (\bibinfo {year} {2011})}\BibitemShut {NoStop}%
\bibitem [{\citenamefont {Luzum}\ and\ \citenamefont
  {Petersen}(2014)}]{Luzum:2013yya}%
  \BibitemOpen
  \bibfield  {author} {\bibinfo {author} {\bibfnamefont {M.}~\bibnamefont
  {Luzum}}\ and\ \bibinfo {author} {\bibfnamefont {H.}~\bibnamefont
  {Petersen}},\ }\href {\doibase 10.1088/0954-3899/41/6/063102} {\bibfield
  {journal} {\bibinfo  {journal} {J. Phys.}\ }\textbf {\bibinfo {volume}
  {G41}},\ \bibinfo {pages} {063102} (\bibinfo {year} {2014})}\BibitemShut
  {NoStop}%
\bibitem [{\citenamefont {Adam}\ \emph {et~al.}(2017)\citenamefont {Adam} \emph
  {et~al.}}]{Adam:2015gka}%
  \BibitemOpen
  \bibfield  {author} {\bibinfo {author} {\bibfnamefont {J.}~\bibnamefont
  {Adam}} \emph {et~al.} (\bibinfo {collaboration} {ALICE}),\ }\href {\doibase
  10.1140/epjc/s10052-016-4571-1} {\bibfield  {journal} {\bibinfo  {journal}
  {Eur. Phys. J.}\ }\textbf {\bibinfo {volume} {C77}},\ \bibinfo {pages} {33}
  (\bibinfo {year} {2017})}\BibitemShut {NoStop}%
\bibitem [{\citenamefont {Aaboud}\ \emph {et~al.}(2016)\citenamefont {Aaboud}
  \emph {et~al.}}]{Aaboud:2016itf}%
  \BibitemOpen
  \bibfield  {author} {\bibinfo {author} {\bibfnamefont {M.}~\bibnamefont
  {Aaboud}} \emph {et~al.} (\bibinfo {collaboration} {ATLAS}),\ }\href
  {\doibase 10.1140/epjc/s10052-016-4335-y} {\bibfield  {journal} {\bibinfo
  {journal} {Eur. Phys. J.}\ }\textbf {\bibinfo {volume} {C76}},\ \bibinfo
  {pages} {502} (\bibinfo {year} {2016})}\BibitemShut {NoStop}%
\bibitem [{\citenamefont {Aamodt}\ \emph
  {et~al.}(2010{\natexlab{b}})\citenamefont {Aamodt} \emph
  {et~al.}}]{Aamodt:2010pp}%
  \BibitemOpen
  \bibfield  {author} {\bibinfo {author} {\bibfnamefont {K.}~\bibnamefont
  {Aamodt}} \emph {et~al.} (\bibinfo {collaboration} {ALICE}),\ }\href
  {\doibase 10.1140/epjc/s10052-010-1350-2} {\bibfield  {journal} {\bibinfo
  {journal} {Eur. Phys. J.}\ }\textbf {\bibinfo {volume} {C68}},\ \bibinfo
  {pages} {345} (\bibinfo {year} {2010}{\natexlab{b}})}\BibitemShut {NoStop}%
\bibitem [{\citenamefont {Breakstone}\ \emph {et~al.}(1984)\citenamefont
  {Breakstone} \emph {et~al.}}]{Breakstone:1983ns}%
  \BibitemOpen
  \bibfield  {author} {\bibinfo {author} {\bibfnamefont {A.}~\bibnamefont
  {Breakstone}} \emph {et~al.} (\bibinfo {collaboration}
  {Ames-Bologna-CERN-Dortmund-Heidelberg-Warsaw}),\ }\href {\doibase
  10.1103/PhysRevD.30.528} {\bibfield  {journal} {\bibinfo  {journal} {Phys.
  Rev.}\ }\textbf {\bibinfo {volume} {D30}},\ \bibinfo {pages} {528} (\bibinfo
  {year} {1984})}\BibitemShut {NoStop}%
\bibitem [{\citenamefont {Petersen}(2011)}]{Petersen:2011sb}%
  \BibitemOpen
  \bibfield  {author} {\bibinfo {author} {\bibfnamefont {H.}~\bibnamefont
  {Petersen}},\ }\href {\doibase 10.1103/PhysRevC.84.034912} {\bibfield
  {journal} {\bibinfo  {journal} {Phys. Rev.}\ }\textbf {\bibinfo {volume}
  {C84}},\ \bibinfo {pages} {034912} (\bibinfo {year} {2011})}\BibitemShut
  {NoStop}%
\end{thebibliography}%
\bibliographystyle{apsrev4-1}
\end{document}